\documentclass[%
reprint,
showpacs,preprintnumbers,
showkeys,
amsmath,amssymb,
aps,
pra,
floatfix,superscriptaddress
]{revtex4-1}

\usepackage[english]{babel}
\usepackage[utf8]{inputenc}
\usepackage{graphicx}
\usepackage{hyperref}
\usepackage{xcolor}
\usepackage{amsmath}
\usepackage{amsfonts}
\usepackage{amssymb}
\usepackage{dsfont}
\usepackage{bm}
\usepackage{todonotes}

\definecolor{refColor}{HTML}{EA00F2}
\definecolor{figColor}{HTML}{008DF2}
\definecolor{urlColor}{HTML}{00AEF2}

\hypersetup{
    unicode=false,                 
    pdftoolbar=true,               
    pdfmenubar=true,               
    pdffitwindow=false,            
    pdfstartview={FitH},           
    pdftitle={},                   
    pdfsubject={Quantum physics},  
    pdfkeywords={},                
    pdfnewwindow=true,             
    colorlinks=true,               
    linkcolor=figColor,            
    citecolor=refColor,            
    filecolor=magenta,             
    urlcolor=urlColor              
}

\newcommand{\bra}[1]{\left\langle #1\right|}             
\newcommand{\ket}[1]{\left| #1\right\rangle}              

\begin{document}

\title{Non-exponential decay of a collective excitation in an atomic ensemble coupled to a one-dimensional waveguide
}

\author{Jan Kumlin}
\affiliation{Institute for Theoretical Physics III and Center for Integrated Quantum Science and Technology, University of Stuttgart, 70550 Stuttgart, Germany}
\author{Kevin Kleinbeck}
\affiliation{Institute for Theoretical Physics III and Center for Integrated Quantum Science and Technology, University of Stuttgart, 70550 Stuttgart, Germany}
\author{Nina Stiesdal}
\affiliation{Department of Physics, Chemistry and Pharmacy, Physics@SDU, University of Southern Denmark, 5320 Odense, Denmark}
\author{Hannes Busche}
\affiliation{Department of Physics, Chemistry and Pharmacy, Physics@SDU, University of Southern Denmark, 5320 Odense, Denmark}
\author{Sebastian Hofferberth}
\affiliation{Department of Physics, Chemistry and Pharmacy, Physics@SDU, University of Southern Denmark, 5320 Odense, Denmark}
\author{Hans Peter B\"uchler}
\affiliation{Institute for Theoretical Physics III and Center for Integrated Quantum Science and Technology, University of Stuttgart, 70550 Stuttgart, Germany}

\date{\today}



\begin{abstract}
We study the dynamics of a single excitation coherently shared amongst an ensemble of atoms and coupled to a one-dimensional wave guide.
The coupling between the matter and the light field gives rise to collective phenomena such as superradiant states with an enhanced initial decay rate, but also to
the coherent exchange of the excitation between the atoms. We find that the competition between the two phenomena provides a characteristic
dynamics for the decay of the excitations, and remarkably exhibits an algebraic behavior, instead of the expected standard exponential one, for a large number of atoms.  The analysis is first performed for a chiral waveguide, where the problem can be solved analytically. Remarkably, we demonstrate that a bidirectional wave guide exhibits the same behavior for large number of atoms, and therefore, it is possible to experimentally access characteristic  properties of a chiral wave guide also within a bidirectional wave guide.
\end{abstract}

\maketitle


\section{Introduction}
\label{sec:introduction}

Coupling light to an ensemble of emitters is one of the paradigmatic models in quantum optics and gives rise to interesting collective and cooperative effects \cite{Guerin2016a}. The most prominent example is superradiance \cite{Dicke1954,Gross1982},
where ensembles of many excited emitters emit at higher intensities if they are excited collectively, rather than independently. Superradiance and other cooperative effects have been observed in a broad spectrum of physical systems ranging from ensembles of nuclei \cite{Rohlsberger2010} over cold atoms \cite{Pellegrino2014,Jennewein2016, Glicenstein2020}, ions \cite{Meir2014}, solid-state systems \cite{Scheibner2007, Tighineanu2016} to more artificial and hybrid light-matter systems like superconducting qubits \cite{mlynek2014observation, PhysRevB.94.224510} or atoms coupled to nanophotonic structures \cite{Goban2015}. Intimately connected to the appearance of superradiant properties of an ensemble is the existence of subradiant states with a strongly quenched emission. These subradiant states find potential applications, for example, in photon storage \cite{Asenjo-Garcia2017} or quantum computing \cite{Petrosyan2002}. However, interesting phenomena appear even in a very weakly excited system with only a single excitation coherently shared among all emitters \cite{Scully2006, Scully2009, Scully2009a, Manassah2009, Mazets2007, Friedberg2008, Mirza2016}. Due to the collective light-matter coupling, for example, the emission rate from the sample is still enhanced compared to an independent emission and scales linearly with the number of emitters. Here, we study the emission dynamics of a single coherent excitation in a superradiant state from an ensemble of emitters coupled to a one-dimensional waveguide.

The influence of collective effects is two-fold. On one hand, the coupling of the ensemble to an external light field is collectively enhanced which can be used to strongly couple a propagating light pulse to an ensemble of many atoms in order to drive Rabi oscillations with only a few photons \cite{Paris-Mandoki2017}. This collective coupling also leads to a strongly enhanced emission rate and the emission becomes highly directional \cite{Scully2006,Guerin2016,Bettles2018}. On the other hand, coherent interactions mediated by the exchange of virtual photons between the emitters were shown to give rise to a collective Lamb shift \cite{Lehmberg1970, Manassah2009}, universal internal dynamics of the ensemble \cite{Kumlin2018} but also strongly influence the decay dynamics of single photon superradiance in three dimensions \cite{Scully2009a, Mazets2007,Svidzinsky2010}. Moreover, coherent interactions can be used to create quantum antennas \cite{Grankin2018}, cavities built from only two atoms \cite{Chang2012} or mirrors built from a single layer of atoms \cite{Bettles2016, Shahmoon2017, Rui2020}. Recently, the efficient coupling of atoms to nanophotonic structures in low dimensions \cite{Vetsch2010} has enabled the study of almost perfectly one-dimensional systems that show infinite-range interactions \cite{Solano2017} but also exotic chiral, coherent light-matter interactions which depend on the polarisation of the incoming light \cite{Lodahl2017}. Such wave guides have a high potential to generate non-classical states of light \cite{Mahmoodian2018, Olmos2020, Iversen2020}.

In this paper, we consider an ensemble of two-level atoms coupled to a one-dimensional waveguide, and study the
emission dynamics of a single excitation coherently shared by all emitters. The approach is based on
the master equation for the atoms describing the coherent interaction by the exchange of virtual photons as well
as the collectively enhanced emission of photons into the waveguide; the master equation is rigorously derived by integrating
out the electric field.  Within this approach, we can distinguish between a chiral waveguide, where atoms only couple to photons propagating
in forward direction as well as a normal waveguide, where  forward and backward propagating photons are treated equally.
The main difference between the two cases appears  in the coherent exchange interaction.  We derive an analytic solution to the master equation describing the dynamics of the
collective single excited state in the chiral waveguide, and find that the probability of having an atomic excitation decays with an algebraic
power law instead of the conventional exponential decay. This behavior is explained by the coherent interactions, which couple
the collective bright state to the many-fold of dark states;  similar phenomena have been predicted recently for numerical and
approximate approaches in three-dimensions \cite{Bettles2018}. Remarkably, we demonstrate that this characteristic algebraic behavior remains present even for the bidirectional waveguide in the limit of large particle number and extended sample size. This observation suggests that some characteristic properties of chiral wave guides are also accessible experimentally in bidirectional wave guides.

This paper is organized as follows:  We start with a general discussion of the coherent exchange interaction and the collective decay and their relation
to the photonic propagator in Section~\ref{sec:model}. We put particular emphasis on one-dimensional waveguides with both chiral and bidirectional coupling and discuss their fundamental difference. In order to illustrate the effect of the coherent exchange on the decay dynamics and understand the underlying process, we examine the simple case of only two atoms coupled to the waveguide in Section~\ref{sec:two_atoms}. Finally, we generalize the model to an arbitrary number of particles in Section~\ref{sec:n_atoms} where an analytical result for the decay of a collective excitation is presented and discussed in view of the understanding gained in the previous section. In addition, we discuss the influence of backscattering in large and small samples.

\section{Model and Results}
\label{sec:model}

\subsection{General approach for the master equation}

We consider a system of $N$ noninteracting two-level atoms at positions $r_j$, where each atom has
a ground state $\ket{g}$ and an excited state $\ket{e}$, separated by the transition
frequency $\omega_0 = c k$. The coupling between the atoms and the electromagnetic field is described within
the rotating frame and applying the rotating wave approximation. The Hamiltonian then takes the form
\begin{equation}
	H = H_0 
	-\hbar \sqrt{\gamma} \sum_{j=1}^N  \left[\mathcal{E}^-(r_j) \sigma_j^+ + \mathcal{E}^+(r_j) \sigma_j^- \right].
	\label{eq:hamiltonian}
\end{equation}
The first term, $H_0$, accounts for the free part of the electromagnetic field and includes the effect of geometric confinement, while the second term accounts for the coupling between the photons and the atoms with strength$\sqrt{\gamma}$.
Here, $\sigma_{j}^{+} = |e\rangle\langle g|_j$ and  $\sigma_{j}^{-}= |g\rangle\langle e|_j$ are
the raising and lowering operators for the atomic transition, while $\mathcal{E}^-$ ($\mathcal{E}^+)$
denotes the positive (negative) frequency component of the electromagnetic field operator; note that the scalar product of
the polarization with the dipole transition moment is included in the definition of  $\mathcal{E}^\pm$.

At any time $t$, the electric field at position $r$ is determined by the radiation field from the spontaneous emission of the atoms and the free field, which account for the quantized field of the incoming photons, \cite{Lehmberg1970, Chang2012, Shi2015, Pichler2015,Ruostekoski2016a, Lodahl2017,LeKien2017}
\begin{equation}
	\mathcal{E}^-(r,t) = \mathcal{E}^-_{\mathrm{free}}(r,t)
		+ \sqrt{\gamma}\sum_{j=1}^N G(r, r_j, \omega_0) \sigma_j^-(t).
\label{eq:electric_field}
\end{equation}
Here, $G(r,r_j,\omega_0)$ is the propagator for the photon field. The precise form
of the propagator is determined by $H_{0}$ and depends on the dimension and geometry of the problem at hand. Note that in Eq.~(\ref{eq:electric_field}), the propagator is local in time. This form is valid if the dispersion relation is well described by a linear behavior around the resonance frequency of the transition for all relevant modes. In addition, retardation effects due to the propagation of photons are neglected. These approximations are usually well satisfied in quantum optical experiments with (cold) atoms and will be discussed in more detail in the next section. With the expression for the quantized electric field, Eq. (\ref{eq:electric_field}), it is then straightforward to derive the master equation describing the atoms alone. Such a derivation has been performed in the past for the general three-dimensional setup  \cite{Lehmberg1970} as well as recently for one-dimensional chiral and non-chiral waveguides \cite{Chang2012, Shi2015, Pichler2015, Ruostekoski2016a, Lodahl2017, LeKien2017}.

Introducing the reduced density matrix $\rho$ describing the atoms alone,  the master equation takes the form \cite{Lehmberg1970} (see also Appendix \ref{app:master_equation} for a more detailed derivation)
\begin{align}
    \notag
    \partial_t \rho(t) &=
    -\frac{i}{\hbar}\left[\hbar \sum_{j,l} J_{jl} \sigma_j^+ \sigma_l^-, \rho(t) \right] \\
    & \qquad + \sum_{j,l} \Gamma_{jl} \left( \sigma_l^- \rho(t) \sigma_j^+
    - \frac{1}{2}\{ \sigma_j^+ \sigma_l^-, \rho(t)\} \right) \, ,
    \label{eq:master_equation}
\end{align}
Here, the first term describes the coherent interaction induced by the exchange of virtual photons, while the last term accounts for the
spontaneous emission. The interaction strengths and decay rates are related to the propagator via
\begin{align}
	J_{jl} &= - \gamma\frac{G^*(r_l, r_j, \omega_0) + G(r_j, r_l, \omega_0)}{2} \, , \\
	\Gamma_{jl} &= i \gamma (G^*(r_l, r_j, \omega_0) - G(r_j, r_l, \omega_0)) \, .
\end{align}
The term $J_{jj}$ accounts for a Lamb shift and is usually dropped as the Lamb shift is already included in the resonance frequency of a single emitter. In turn, $\Gamma_{jj}$ describes the single-emitter decay rate. Note that the above expressions are general and do not assume any symmetry of $G$ itself. This becomes crucial
when we consider a one-dimensional chiral waveguide in which the propagator is not symmetric under exchange of two atoms.

\subsection{One-dimensional waveguide}

\begin{figure}
    \centering
    \includegraphics[width=0.45\textwidth]{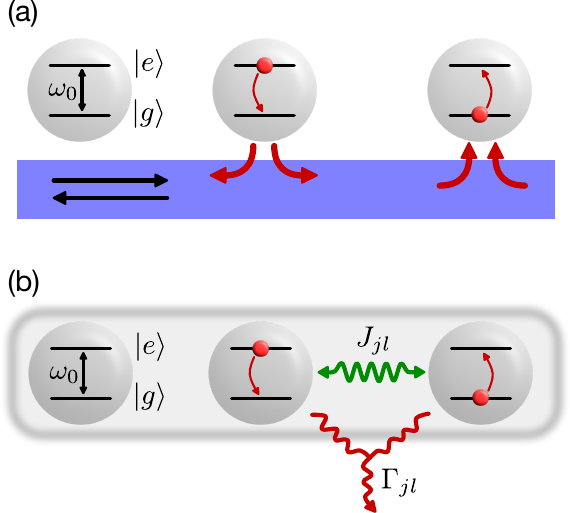}
    \caption{(a) Two-level atoms coupled to a one-dimensional waveguide. The waveguide supports (in general) left- and right-propagating modes and the atoms can emit (absorb) photons into (from) both modes. (b) Effective system after the elimination of the waveguide photons. The atoms interact via an (infinite-ranged) exchange interaction $J_{jl}$ and have a correlated decay $\Gamma_{jl}$.}
    \label{fig:setup_general}
\end{figure}

For the remainder of this paper, we focus on one-dimensional waveguides (see Fig.~\ref{fig:setup_general}). In particular, we are interested in chiral waveguides, where the atoms are only coupled to forward-propagating modes, and bidirectional waveguides, where the atoms couple to both forward- and backward-propagating modes. In addition, our focus is on optical setups with transition frequencies $\omega_0$ on the order of a few hundred THz and coupling constants $\gamma$ in the MHz regime. Since these time scales are well separated, the rotating-wave approximation performed in (\ref{eq:hamiltonian}) is justified.

In the following, we outline the derivation of the propagator of the photon field for a one-dimensional waveguide while more details can be found in Appendix \ref{app:photon_propagator}. First, we focus on the chiral waveguide, where the Hamiltonian for the photons in the rotating frame of the atoms takes the form
\begin{align}
    H_0 = \int_{k-q_c}^{k+q_c} \frac{dq}{2\pi} \hbar \omega_q a_q^\dagger a_q \, .
\end{align}
Here, $\omega_q = c q - \omega_0$ denotes the dispersion relation for the photons, which is assumed to be linear around the resonance frequency $\omega_0$ of the atoms. The bosonic operator $a_q^\dagger$ ($a_q$) describes the creation (annihilation) of a photon with momentum $q$. We have introduced a cut-off parameter $q_c$, which accounts for the momentum regime, where the description of the dispersion relation by a linear spectrum is valid. In the experimentally relevant regime with $N \gamma \ll c q_c \ll ck$, the cut-off can be removed in the derivation of the master equation; see below.

Then, the electric field operator is given by
\begin{align}
    \mathcal{E}^-(x) = i \sqrt{c} \int_{k - q_c}^{k+q_c} \frac{dq}{2\pi} a_q e^{i q x} \, .
    \label{eq:e-field_1D}
\end{align}
This allows us now to derive the propagator of the photon field in Eq.(\ref{eq:electric_field}) for the chiral waveguide. We start with the
 time evolution of the electric field operator, which is obtained by formally integrating the Heisenberg equation of motion $i \hbar \partial_t a_q = [a_q,H]$,
\begin{align}
    a_q(t) = & \: e^{- i \omega_q t} a_q(0) \\ &+ \sqrt{\gamma\, c} \sum_j \int_0^t ds\, e^{- i q x_j} e^{- i \omega_q (t-s)} \sigma_j^-(s) \, .   \nonumber
\end{align}
Plugging this expression into Eq.~(\ref{eq:e-field_1D}), leads to the input-output relation
\begin{align}
    \mathcal{E}^-(x,t) = & \: \mathcal{E}^-_\text{free}(x,t) + i \sqrt{\gamma} \sum_j \int_0^t ds  \\
    & \quad \times\int_{\omega_0 - \omega_c}^{\omega_0 + \omega_c} \frac{d\omega}{2\pi} e^{i \frac{\omega}{c}(x- x_j)} e^{- i (\omega- \omega_0)(t-s)} \sigma_j^-(s) \, .  \nonumber
\end{align}
The first term in this expression corresponds to the non-interacting part of the electric field, while the latter one describes the interaction with the emitters. Note that we converted the integral over momentum $q$ into an integral over frequency $\omega$ and that $\omega_c = c q_c$ is a cutoff frequency.

In order to derive the propagator of the photon field (\ref{eq:electric_field}), we perform the narrow-bandwidth approximation assuming that the atomic operators $\sigma_j^-$ vary only slowly on a time scale $ N \gamma$ and that $N \gamma \ll \omega_c \ll \omega_0$. The frequency integration can then be replaced by a $\delta$-function at the retarded time $t - (x-x_j)/c$ as long as $x \geq x_j$ such that the electric field takes the form
\begin{align}
    \mathcal{E}^-(x,t) = & \: \mathcal{E}^-_\text{free}(x,t) \\
    & + i \sqrt{\gamma} \sum_j \theta(x - x_j) e^{i k (x - x_j)} \sigma_j^-\left(t - \frac{x - x_j}{c} \right) \, ,  \nonumber
\end{align}
where $\theta(x)$ is the Heaviside function with $\theta(0) = 1/2$ and $k = \omega_0/c$.

For typical quantum optical systems as discussed above, we have $N \gamma \ll c/\vert x - x_j \vert$, which allows us to neglect the influence of retardation within the waveguide, i.e., the time scale for the propagation of a photon from one atom to the next is short compared to the characteristic dynamics. For a realistic system with coupling strengths on the order of a few MHz and atom number of a few thousand to ten thousand atoms, this leads to an maximum width of the sample on the order of a few centimeters to millimeters depending on the precise number of atoms and coupling strengths. This is well beyond the length scales of micrometers of those systems of interest. Then, we can approximate $\sigma_j^-(t - (x-x_j)/c) \approx \sigma_j^-(t)$. Comparison with (\ref{eq:electric_field}) leads to the propagator for a one-dimensional chiral waveguide, which reads
\begin{equation}
    G_{\rm \scriptscriptstyle chiral}(x_j, x_l) = i e^{i k (x_j - x_l)} \theta(x_j - x_l)\, .
\end{equation}
This propagator is not symmetric under particle exchange, that is $G_{\rm \scriptscriptstyle chiral}(x_j, x_l) \neq G_{\rm \scriptscriptstyle chiral}(x_l, x_j)$. The coherent exchange terms and decay rates read
\begin{align}
    J_{jl} &= \frac{\gamma}{2i} \text{sign}(x_j - x_l) e^{i k (x_j - x_l)} \, , \\
    \Gamma_{jl} & = \gamma e^{ i k (x_j - x_l)} \, ,
\end{align}
where the single-atoms decay rate is $\Gamma_{jj} = \gamma$.

If we turn to a bidirectional waveguide, where the atoms are coupled to modes with positive and negative momenta, we can perform a similar calculation using the same approximations. This results in the propagator for the bidirectional waveguide given by
\begin{align}
    G(x_j, x_l) = i e^{i k \vert x_j - x_l \vert} \, ,
\end{align}
which exhibits the symmetry $G(x) = G(-x)$.
The corresponding coherent exchange terms and decay rates read
\begin{align}
    J_{jl}  &= \gamma \sin( k \vert x_j - x_l \vert) \, ,
    \label{eq:exchange_1D_bidirectional} \\
    \Gamma_{jl}  &= 2 \gamma \cos( k \vert x_j - x_l \vert) \, ,
    \label{eq:decay_1D_bidirectional}
\end{align}
with the single-atom decay rate $\Gamma_{jj} = 2 \gamma$. Note, that the single atom decay rate is twice as large for the bidirectional waveguide, as the photon can be emitted into the forward and the backward propagating mode.

\section{Two-atom solution}
\label{sec:two_atoms}

\begin{figure}
    \centering
    \includegraphics[width = 0.47\textwidth]{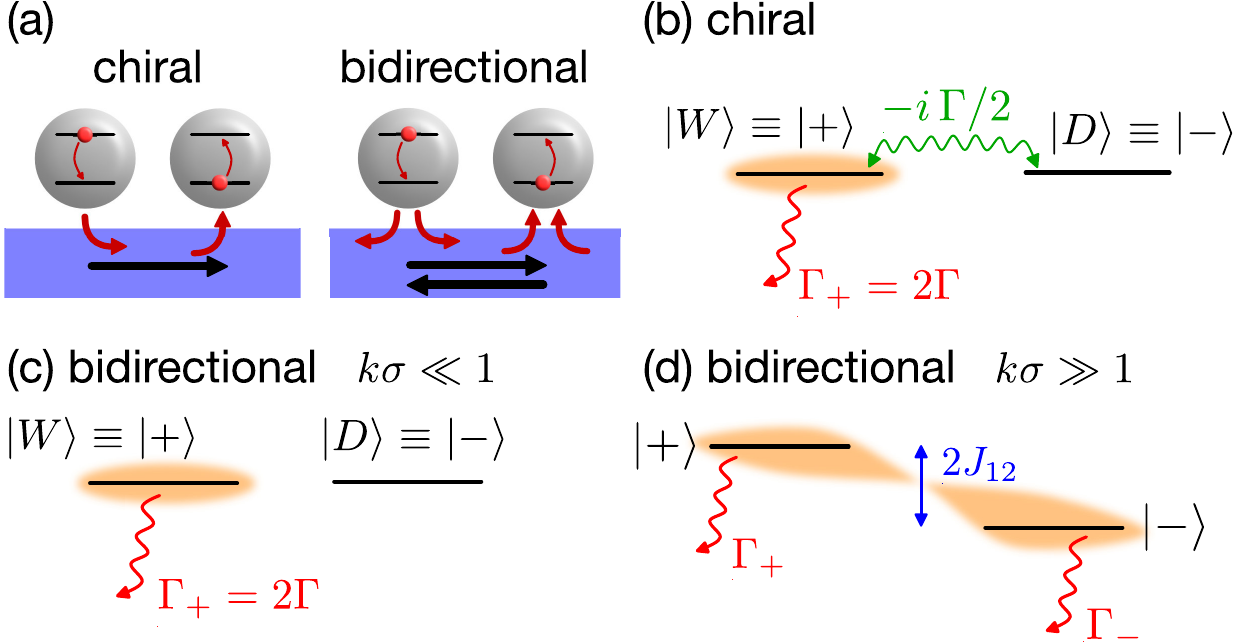}
    \caption{Setup for two atoms coupled by a one-dimensional waveguide. (a) In a chiral setup (left) the atom can only emit into the forward-propagating mode with rate $\gamma$, while in the bidirectional setup (right) the atoms can emit into forward- and backward-propagating mode with rate $\gamma$ for each mode. (b) For a chiral waveguide, the subradiant and superradiant states correspond to the bright and dark state, respectively. The bright and dark state are coupled and the bright state decays with a collectively enhanced decay rate $\Gamma_+ = 2 \Gamma$ and the single-atom emission rate $\Gamma = \gamma$. Initially, the system is prepared in the bright state. (c) In the bidirectional case, when the atoms are on average very close to each other compared to the wavelength, there is no coupling and the bright (dark) state corresponds to the superradiant (subradiant) state. A system that is initially prepared in the bright state decays with the collectively enhanced decay rate $2 \Gamma$, where the single-atom decay rate is $\Gamma = 2 \gamma$. (d) In the case where the interatomic distance is on average much greater than the wavelength, the superradiant and subradiant state are shifted with respect to each other (depending on the distance between the atoms) and emit with rates $\Gamma_+$ and $\Gamma_-$, respectively. Since the bright state is now a superposition of super- and subradiant state, these two states are coupled by the initial condition.}
    \label{fig:setup_2atoms}
\end{figure}

As an illustrative example that already contains the important physics, we review the case of only two atoms, which has also been studied extensively in previous works \cite{Berman2020,Jen2020}. Consider the generic master equation for a system of only two identical atoms at positions $x_1$ and $x_2$ given by

\begin{align}
\partial_t \rho & = -i  \left[ J_{12} \sigma_1^+ \sigma_2^- + J_{12}^* \sigma_2^+ \sigma_1^-, \rho\right] \nonumber \\
& \quad + \Gamma \bigg( \mathcal{D}[\sigma^-_1] \rho + \mathcal{D}[\sigma^-_2] \rho \nonumber \\
& \quad + F_{12} \left(\sigma^-_2 \rho \sigma^+_1 - \frac{1}{2} \{\sigma_1^+ \sigma_2^-, \rho \}\right) \nonumber \\
& \quad + F^*_{12} \left( \sigma_1^- \rho \sigma_2^+ - \frac{1}{2}\{ \sigma_2^+ \sigma_1^-, \rho\}\right) \bigg) \, ,
\label{eq:ME2_1}
\end{align}
where $J_{12} \in \mathbb{C}$ is the coherent coupling between the atoms, $\mathcal{D}[\sigma^-]\rho = \sigma^- \rho \sigma^+ - 1/2\{ \sigma^+ \sigma^-, \rho \}$ is the Lindblad dissipator and $\Gamma$ is the single-atom emission rate into the waveguide. In a chiral waveguide there is only a coupling to the forward propagating modes and the single-atoms emission rate is $\Gamma = \gamma$, whereas for a bidirectional waveguide, the atom can emit into forward- and backward-propagating modes and the emission rate is $\Gamma = 2\gamma$ (see Fig.~\ref{fig:setup_2atoms}a).

The dimensionless factor $F_{12} \in \mathbb{C}$ is a measure for the correlated decay of both atoms in terms of $\Gamma$.  If $F_{12} = 0$, the atoms decay independently of each other with the single-atom decay rate $\Gamma$. If $F_{12}$ is different from zero, the decay rates are modified in general and in the single-excitation subspace, there is one superradiant state which decays faster than $\Gamma$ and one subradiant state which decays slower than $\Gamma$. The super- and subradiant states read
\begin{align}
\ket{\pm} = \frac{1}{\sqrt{2}} \left( \sigma_1^+ \pm e^{-i \phi} \sigma_2^+  \right) \ket{G} \equiv \frac{1}{\sqrt{2}} S_{\pm}^\dagger \ket{G}
\label{eq:super_subradiant_2}
\end{align}
where $\phi = \text{arg}(F_{12})$ and $\ket{G}$ is the ground state of the atomic system, where all atoms are in their respective ground state. The corresponding decay rates are $\Gamma_{\pm} = \Gamma(1 \pm \vert F_{12} \vert)$. Note that the decay rates depend on the distance between the emitters. While the super- and subradiant states provide an elegant way to describe the decay dynamics of a single excitation, for actual experiments another type of state is of importance. Assume that in a one-dimensional setup the system is excited by means of a plane wave $e^{ikx}$. In the single-excitation sector the light field couples to the so-called bright state
\begin{equation}
    \ket{W} = \frac{1}{\sqrt{2}} (\sigma_1^+ + e^{- i k (x_1 - x_2)}\sigma_2^+) \ket{G} \equiv \frac{1}{\sqrt{2}}S_W^\dagger \ket{G}\, .
\end{equation}
The orthogonal state
\begin{equation}
    \ket{D} = \frac{1}{\sqrt{2}} ( \sigma_1^+ - e^{- i k (x_1 - x_2)} \sigma_2^+) \ket{G} \equiv \frac{1}{\sqrt{2}}S_D^\dagger \ket{G}\,
\end{equation}
is called the dark state and is decoupled from the incoming light field. It is important to note that while the bright and dark state look similar to the super- and subradiant state defined in Eq.(\ref{eq:super_subradiant_2}), they coincide only in very special cases as we will show in the following.

\subsection{Bidirectional waveguide}

First, we focus on the bidirectional waveguide for which $\Gamma = 2 \gamma$, $F_{12} = \cos(k \vert x_1 - x_2 \vert)$ and $J_{12} = \frac{\Gamma}{2} \sin(k \vert x_1 - x_2 \vert) \in \mathbb{R}$, which can be inferred by comparing eqs. (\ref{eq:exchange_1D_bidirectional}), (\ref{eq:decay_1D_bidirectional}) and (\ref{eq:ME2_1}). The resulting master equation for this system reads
\begin{align}
    \partial_t \rho & = -i \left[ J_{12} ( S_+^\dagger S_+ - S^\dagger_- S_-), \rho \right] \nonumber \\
    & {} \quad + \Gamma_+ \mathcal{D}[S_+] \rho + \Gamma_- \mathcal{D}[S_-]\rho \, .
\end{align}
Note that the dynamics for the super- and subradiant states completely decouple, and both states are shifted by $J_{12}$ with respect to each other. This situation is qualitatively similar to a system of two atoms coupled to the electromagnetic continuum in free space as the parameters $J_{12}$ and $F_{12}$ are real and depend on the relative distance between the atoms. The precise form of the coupling parameter and decay rates, however, are much more complicated and also depend on the relative orientation of the two atoms.

The dynamics of the system of two atoms can be calculated analytically for arbitrary positions of the atoms and by defining the elements of the density matrix $\rho_{\alpha \beta} = \bra{\alpha} \rho \ket{\beta}$. The populations of the bright state and dark state for a system initially prepared in the bright state are given by
\begin{align}
\rho_{WW}(t) & = e^{- \Gamma t} \left\vert \cosh\left(\frac{\Gamma t}{2}  \, e^{i k \vert x_1 - x_2 \vert}\right) \right. \nonumber \\
& \left. \qquad - \cos( k \, (x_1 - x_2)) \sinh\left(\frac{\Gamma t}{2} \, e^{i k \vert x_1 - x_2 \vert} \right) \right\vert^2\, , \label{eq:rhoww_bidirectional}\\
\rho_{DD}(t) & = e^{- \Gamma t} \sin^2(k \vert x_1 - x_2\vert) \left\vert \sinh\left(\frac{\Gamma t}{2} \, e^{i k \vert x_1 - x_2 \vert}\right) \right\vert^2\, .
\label{eq:rhodd_bidirectional}
\end{align}

For short distances, $k \vert x_1 - x_2 \vert \ll 1$, one can approximate $F_{12} \approx 1$ and $J_{12} \approx 0$ resulting in $\Gamma_+ = 2 \Gamma$ and $\Gamma_- = 0$. In addition, the bright and dark state coincide with the super- and subradiant state, respectively (see Fig.~\ref{fig:setup_2atoms}c ). In this scenario, the bright state decays exponentially with an enhanced decay rate $2 \Gamma$ known as single-photon superradiance which was already studied by Dicke \cite{Dicke1954}. The same holds when we go to the experimentally more relavant case where the positions of the atoms might fluctuate for different realizations of the experiment. Assuming that the atoms are distributed according to a density distribution with characteristic length scale $\sigma$, single-photon superradiance is also present if $k \sigma \ll 1$, that is if the atoms are much closer than a wavelength. This can be also seen from Eq.~(\ref{eq:rhoww_bidirectional}), which reduces to $\rho_{WW}(t) \approx e^{- 2 \Gamma t}$ in these cases.

\begin{figure}
    \centering
    \includegraphics[width = 0.45\textwidth]{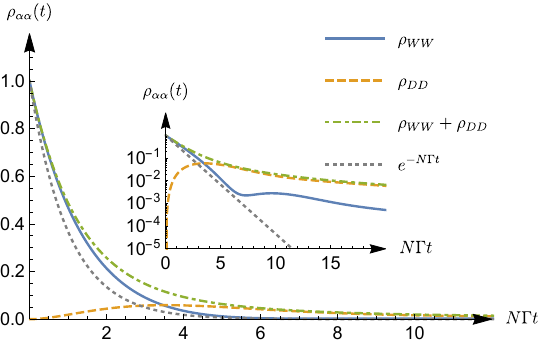}
    \caption{(Color online) Time evolution of the population of the bright state (blue solid line), the dark state (orange dashed line) and the total population of excited states (green dash-dotted line) for $N = 2$ atoms coupled to a bidirectional waveguide in the limit $k \sigma \gg 1$. The grey dotted line shows an exponential decay with a collectively enhanced decay rate $N \Gamma = 2 N \gamma$ expected in single-photon superradiance which appears for $k \sigma \ll 1$. (Inset): The inset shows the time evolution on a logarithmic scale. For small times $N \Gamma t \ll 1$, the decay can be approximated as $1 - \frac{3}{2} N \Gamma t \approx e^{- \frac{3}{2} N \Gamma t}$. For long times $N \Gamma t \gg 1$, the populations decay much slower compared to an exponential decay with collectively enhanced decay rate $ N \Gamma$. The numerical calculations were performed for $k \sigma = 1000$ and the positions of the atoms varied according to a Gaussian density distribution with mean $0$ and variance $\sigma^2$. The plot shows the average over $M = 1000$ realizations and convergence with respect to $M$ was checked.}
    \label{fig:PW_2atoms_bidirectional}
\end{figure}

In the opposite limit where the extent of the ensemble is much larger than the wavelength, that is $k \sigma \gg 1$, the behaviour for small times $\Gamma t \ll 1$ after averaging over the atomic distribution \cite{comment} is
\begin{align}
    \rho_{WW}(t) &\approx 1 - \frac{3}{2}\Gamma t + \mathcal{O}((\Gamma t)^2) \approx e^{- \frac{3}{2}\Gamma t} \, , \\
    \rho_{DD}(t) &\approx \frac{1}{8}(\Gamma t)^2 + \mathcal{O}((\Gamma t)^3) \, .
\end{align}
Thus, the bright state initially does not decay with a collectively enhanced rate $2 \Gamma$, but slightly slower due to the additional decay channel in the backward direction. The full, numerical solution for the time evolution of the bright state, the dark state and the overall population of the excited states, $\rho_{WW} + \rho_{DD}$, is shown in Fig.~\ref{fig:PW_2atoms_bidirectional} alongside with the time evolution for the superradiant case. It can be seen that for longer times, $\Gamma t \gg 1$ the population of the bright state together with the overall population of the excited states decay much slower than expected from a superradiant sample due to the influence of the dipole-dipole interactions.

\subsection{Chiral waveguide}

Next, we study a chiral waveguide, where each atom only couples to the forward propagating modes of the waveguide and the emission of each atom is directional with rate $\Gamma = \gamma$. The measure of the collective decay is $F_{12} = e^{i k (x_1 - x_2)}$ and carries the phase the photon picks up when propagating from one atom to the other. The exchange coupling parameter reads $J_{12} = \frac{\Gamma}{2 i} \text{sign}(x_1 - x_2) e^{i k (x_1 - x_2)}$ and is in general complex. As the correlated decay term, the exchange coupling also carries the phase of the photon due to propagation while the $\text{sign}$ term comes from the chiral coupling.

\begin{figure}
    \centering
    \includegraphics[width = 0.45\textwidth]{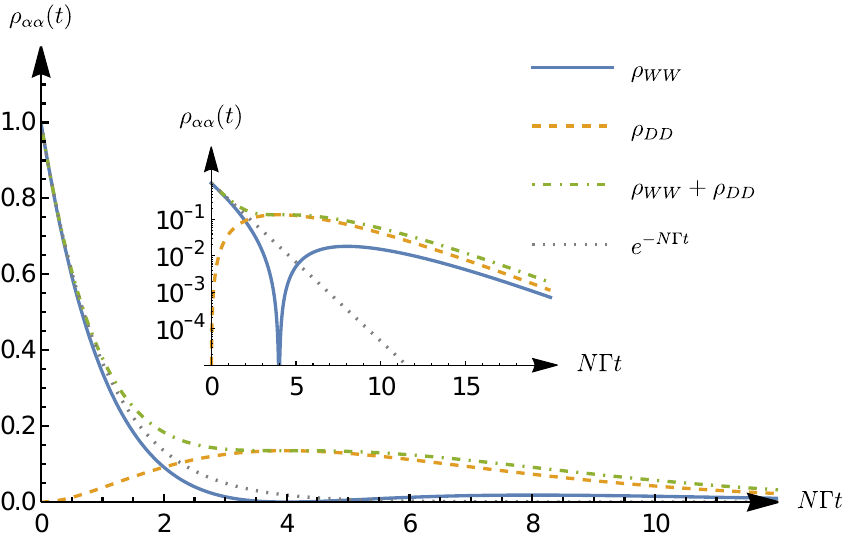}
    \caption{(Color online) Time evolution of the population of the bright state (blue solid line), the dark state (orange dashed line) and the total population of the excited states (green dash-dotted line) for $N = 2$ atoms coupled to a chiral waveguide. The gray dotted line shows an exponential decay with collectively enhanced decay rate $N \Gamma$. (Inset): The inset shows a logarithmic plot of the time evolution of the populations. For $N \Gamma t \gg 1$, the decay of the bright state population is slowed down due to the coupling to the dark state.}
    \label{fig:PW_2atoms_chiral}
\end{figure}

For the chiral system, the definition of the bright (dark) state $\ket{W}$ ($\ket{D}$) conincides with the definition of the superradiant (subradiant) state $\ket{+}$ ($\ket{-}$), see also Fig.~\ref{fig:setup_2atoms}b). As a matter of fact, neither the precise positions of the atoms nor their relative distance matter for the physics but only their ordering with respect to each other. This is due to the cascaded nature of the system, where the atoms can only emit into the forward direction, which coincides with the direction of propagation of the incoming plance wave. This can also be seen by redefinition of the spin operators to include the propagation phase, i.e. $e^{- i k (x_1 - x_2)}\sigma_2^+ \to \sigma_2^+$. The master equation (\ref{eq:ME2_1}) for the chiral system expressed in terms of super- and subradiant operators reads
\begin{align}
    \partial_t \rho &= - i \left[i \frac{\Gamma}{4}(S_+^\dagger S_- - S_-^\dagger S_+ ) , \rho \right] + \Gamma_+ \mathcal{D}[S_+] \rho \nonumber \\
    & = - i \left[i \frac{\Gamma}{4}(S_W^\dagger S_D - S_D^\dagger S_W) , \rho \right] + \Gamma_+ \mathcal{D}[S_W] \rho\, ,
\end{align}
where $\Gamma_+ = 2 \Gamma = 2 \gamma$ and $\Gamma_- = 0$. This means, that the super- and subradiant state in this case are perfectly superradiant and subradiant, respectively. In addition, we have assumed $x_1 < x_2$ for simplicity. In contrast to the bidirectional case, the master equation does not decouple into super- and subradiant states but coherently couples them due to the chiral coupling (see also Fig.~\ref{fig:setup_2atoms}). Preparing the system in the bright state, which is equivalent to the superradiant state, the bright state can either decay with enhanced rate $\Gamma_+ = 2 \Gamma$ or couple to the dark (subradiant) state that does not decay at all. Since the coupling is a coherent process, the system will decay with $\Gamma_+$ in linear order. For later times, the probability to remain in the bright state will no longer follow an exponential decay with enhanced decay rate $\Gamma_+$ but should first decay faster due to an additional channel to the dark state with a subsequent revival due to coupling back from the dark state. The time evolution for the population of the bright and dark state of a system initially prepared in the bright state reads
\begin{align}
    \rho_{WW}(t) &= \frac{1}{4}e^{-\Gamma t}(\Gamma t - 2)^2 \, , \\
    \rho_{DD}(t) &= \frac{1}{4} e^{-\Gamma t} (\Gamma t)^2 \,
\end{align}
and is also shown in Fig.~\ref{fig:PW_2atoms_chiral}. As discussed before, for short times, $\Gamma t \ll 1$, the bright state decays as $\rho_{WW}(t) \approx 1- 2 \Gamma t \approx e^{- 2 \Gamma t}$, while it vanishes for $\Gamma t = 2$, will have a revival shortly after and then decays again. The rapid decrease of the population of the bright state after some initial time must not be confused with the spontaneous emission of a photon but rather with the transfer of the excitation into the dark state. This can also be seen looking at the corresponding population of the dark state and the probability to find an excitation in the system, given by $\rho_{WW} + \rho_{DD}$. At $\Gamma t = 2$, all population that has not yet decayed is transferred to the dark state. For longer times the decay is not exponential with a collectively enhanced decay rate but rather slows down due to the coupling of the bright, superradiant state to the dark, subradiant one.

\section{$N$ atoms coupled to a one-dimensional waveguide}
\label{sec:n_atoms}

After having shown that including the coherent exchange interaction into the dynamics of a two-emitter system can alter the characteristics of the decay dynamics of a single collective excitation, we proceed to a more complex situation where an arbitrary number $N$ of emitters are coupled to a one-dimensional waveguide. At first glance, it is not obvious that we can expect similar dynamics as in the case of only two emitters as we are dealing with many dark states that are also coupled to each other leaving the possibility of an ordinary exponential decay albeit with a modified decay rate. In the following, we show both numerically and analytically that this is not the case but instead there are oscillations in the population of the bright state with an overall algebraic decay. First, we discuss the case of a chiral coupling, meaning that the photons emitted from the atoms into the waveguide can only propagate into one direction, for example from left to right. Owing to the chiral coupling, it is possible to derive an analytical expression for the population of the bright state. In a second step, we include the emission into the other direction and show that for an extended sample of atoms, the dynamics reduces to that of a chiral waveguide.

\subsection{Chiral waveguide}

Since for a chiral setup the atoms can only emit into one direction, say to the right, they form a cascaded open quantum system \cite{Gardiner1993, Carmichael1993}. The corresponding master equation reads \cite{Stannigel2010, Pichler2015, Lodahl2017}
\begin{align}
\partial_t \rho = &- \frac{i}{\hbar} \left[\frac{\hbar \gamma}{2i} \sum_{j,l} \text{sign}(x_j - x_l) e^{ i k (x_j - x_l)} \sigma_j^+ \sigma_l^-, \rho \right] \nonumber \\
& + \gamma \sum_{j,l} e^{ i k (x_j - x_l)}\left( \sigma_l^- \rho \sigma_j^+ - \frac{1}{2} \left\lbrace \sigma_j^+ \sigma_l^- , \rho \right\rbrace \right)\, ,
\end{align}
where $\text{sign}(x-y) = \mp 1$ if $x \lessgtr y$ and $\text{sign}(x-y) = 0$ if $x = y$.
 Again, the specific positions $x_{i}$ of the atoms do not influence the dynamics as the phase factors could be absorbed into the definition of the operators $\sigma^\pm_{i}$.

\begin{figure}
\includegraphics[width=0.45\textwidth]{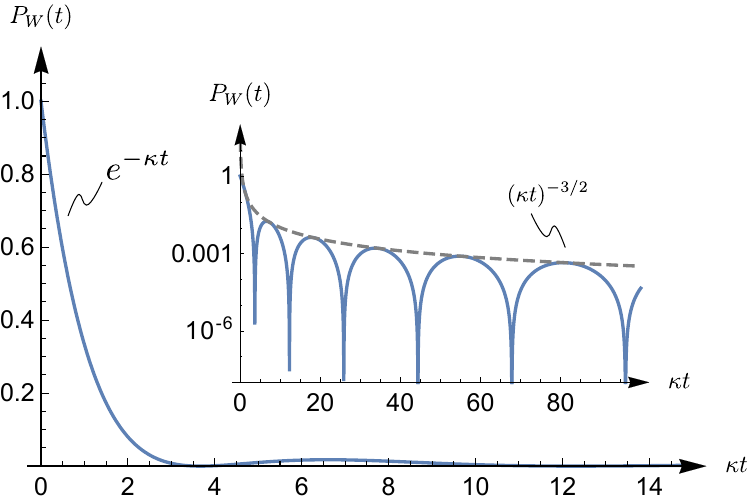}
\caption{Decay dynamics of a single collective excitation of a system of $N$ atoms coupled to a chiral waveguide in the limit $N \to \infty$. The collective excitation initially decays exponentially with decay rate $\kappa$ while for long times the decay is algebraic with $(\kappa t)^{-3/2}$ which is shown in the inset. The dashed line shows the long-time behavior. Note that the dynamics looks qualitatively the same for finite $N$ and $N \gg 1$.}
\label{fig:pw_Natoms}
\end{figure}

Since we are only interested in the dynamics of a single excitation, the time evolution of the system is well described by the effective non-Hermitian Hamiltonian
\begin{equation}
    H_\text{eff} = \frac{\hbar \gamma}{2 i} \sum_{j,l} \left(\text{sign}(x_j-x_l) + 1 \right)e^{i k (x_j - x_l)} \sigma_j^+ \sigma_l^-
\end{equation}
which includes both the coherent exchange coupling and the collective chiral decay. This description is possible since we do not have any external driving and do not assume initial coherences between the single-excitation subspace and the ground state. In what follows, we focus on the modification of the collectively enhanced decay of the state
\begin{equation}
    \ket{W} = \frac{1}{\sqrt{N}} \sum_{j} e^{i k x_j} \sigma_j^+ \ket{G}
\end{equation}
due to the chiral coupling, we consider the quantity
\begin{align}
    P_W(t) = \vert \bra{W} e^{-i H_\text{eff} t / \hbar} \ket{W} \vert^2\, ,
\end{align}
which is identical to the population of the state $\ket{W}$. The time evolution of the $\ket{W}$ state can be calculated analytically for the chiral case and the solution reads (see Appendix \ref{app:analytical} for more details)
\begin{equation}
    P_W(t) = \frac{1}{N^2}e^{-\gamma t} \left[L_{N-1}^{(1)} (\gamma t) \right]^2 \, ,
    \label{eq:brightStateDecay}
\end{equation}
where $L_m^{(n)}(x)$ is the generalized Laguerre polynomial. The decay dynamics of the state $\ket{W}$ are shown in  Fig.~\ref{fig:pw_Natoms}.

For short times, $\gamma t \ll 1$, one expects the coherent exchange to play no role such that the decay is completely determined by the collective decay given by $N \gamma$. Indeed, for short times $N\gamma t \ll 1$, we find
\begin{equation}
P_W( t \ll 1/N\gamma) \approx 1- N \gamma t + \mathcal{O}((N\gamma t)^2) \approx e^{- N \gamma t}\, .
\end{equation}

Eq.(\ref{eq:brightStateDecay}) can be further simplified in the asymptotic limit
$N \to \infty$, and we obtain
\begin{equation}
P_W(t) = \frac{(J_1(2\sqrt{\kappa t}))^2}{\kappa t} \,
\label{eq:pw_asymptotic}
\end{equation}
where $J_n(x)$ is the Bessel function of the first kind and $\kappa = N \gamma$ with $\kappa$
fixed for $N \to \infty$. In the limit $N \to \infty$, the initial decay for short times is
given by $\kappa$, while for long times $\kappa t \gg 1$, we find a characteristic algebraic
behavior
\begin{equation}
P_W(\kappa t \gg 1) = \frac{1}{\pi (\kappa t)^{3/2}} \cos^2\left( 2 \sqrt{\kappa t} - \frac{3 \pi }{4}\right) \, .
\label{eq:N_atom_longtime}
\end{equation}
Interestingly, there is no exponential decay for long times but rather an algebraic one with
$(\kappa t)^{-3/2}$. This is also shown in the inset of Fig.~\ref{fig:pw_Natoms}. For finite $N$
the algebraic decay is present on intermediate timescales $\kappa t \gg 1$. However, the decay of
individual atoms eventually becomes the dominant contribution, which happens on timescales
$\kappa t \gg N^2$. This is in stark contrast to the collectively enhanced exponential decay
one encounters in single-photon superradiance.

The slowing down of the emission from the bright state can be understood as follows: Also for many atoms, the single-excitation subspace can be divided into a superradiant state and subradiant states. The interaction mediated by photon exchange via the waveguide couples the bright superradiant state to the other subradiant states. Therefore, these subradiant states become populated during the time evolution, and the excitation is less likely to decay if it is "protected" in these subradiant states. This mechanism then provides the slowing down of the decay dynamics.

We want to point out that in the limit of $N \to \infty$, the rotating-wave approximation breaks down and neglecting retardation effects is also no longer justified. As a physically meaningful limit, we require always $\kappa = N \gamma \ll \omega_0$. Typical experiments with ultracold atoms, for example, involve about $10^3$ to $10^4$ atoms with coupling constant $\gamma$ in the MHz regime and optical transition frequencies in the THz regime. The above condition is thus well satisfied.

Lastly, we note that a similar study with atoms at fixed positions and slightly asymmetric coupling was performed in \cite{Jen2020}.

\subsection{Bidirectional waveguide: large and small samples}

\begin{figure}
\includegraphics[width=0.45\textwidth]{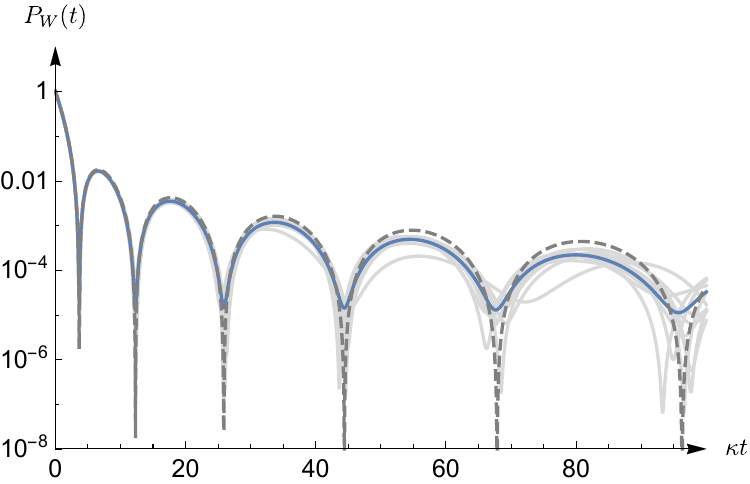}
\caption{(Color online) The blue (dark grey) line shows the time evolution of the bright state in the case of a bidirectional coupling and a normal distribution of the atoms with zero mean and variance $\sigma^2$ with $k \sigma =1000$ for $N = 100$ atoms and averaged over $M = 100$ realizations. The gray dashed line shows the corresponding time evolution for the chiral case for the same number of particles. The light gray curves in the background show trajectories for single realizations.}
\label{fig:pw_bidirectional}
\end{figure}

While we have shown above that the dynamics of a single collective excitation in a one-dimensional chiral waveguide undergoes interesting dynamics, we now turn to the case where the waveguide is bidirectional but the positions of the atoms  fluctuate with each realization. It turns out that in the limit where the distribution of the position of the atoms is smooth compared to the wavelength, one recovers the dynamics of a chiral waveguide. In contrast, the case of an ensemble that is confined within a wavelength shows single-photon superradiance with an collectively enhanced exponential decay of the collective excitation.
First, we consider the case where the atoms are randomly distributed along the waveguide with a characteristic length scale $\sigma$, which is much larger than the wavelength of the atomic transition, that is $k \sigma \gg 1$. The time evolution of the bright state for $N = 100$ atoms  can be determined numerically and is shown in Fig.~\ref{fig:pw_bidirectional}. For concreteness, we use a Gaussian density distribution with width $\sigma$ and $k \sigma = 1000$, the result is averaged over 100 realizations. Interestingly, the dynamics in the bidirectional case are qualitatively similar to the chiral case after averaging over the position of the atoms. Even for single realizations of the system the time evolution of the bidirectional case resembles the dynamics of the chiral system in terms of algebraic decay and period of the oscillations.

In order to understand this observation,
 we can go to the continuum limit for $N \to \infty $, and introduce again the effective Hamiltonian
\begin{equation}
     H_\text{eff} = - i \gamma \int dx \, dy \, \exp(i k \vert x - y \vert) \Psi^\dagger(x) \Psi(y)
\end{equation}
with the field creation and annihilation operators $\Psi^\dagger(x)$ and $\Psi(x)$, respectively. Their commutation relations are $[\Psi(x), \Psi^\dagger(y)] = \delta(x-y)$. The time evolution of the state
\begin{equation}
    \ket{\psi(t)} = \frac{1}{\sqrt{N}} \int dx \, \psi(x,t) \Psi^\dagger(x) \ket{G}
\end{equation}
with the initial condition $\psi(x,0) = e^{i k x}$ is then given by the effective Schr\"odinger equation
\begin{equation}
    \partial_t \psi(x,t) = - \gamma \int dy \, \exp(i k \vert x - y \vert) \psi(y,t) n(y) \, ,
    \label{eq:effective_SE}
\end{equation}
where $n(y)$ is the density distribution of the atoms with a characteristic width $\sigma$ with $\int dx \, n(x) = N$. In the limit $k \sigma \to \infty$ and assuming that the atoms are uniformly distributed in the interval $[0, \sigma]$, this equation can be solved using the Laplace transform with respect to both $t$ and $x$. The solution for $\psi(x,t)$ is given by (for more details, see Appendix \ref{app:continuum})
\begin{equation}
    \psi(x,t) = e^{i k x} J_0( 2 \sqrt{\kappa t x/\sigma})\, .
\end{equation}
The population of the bright state is then given by
\begin{equation}
    P_W(t) = \left\vert \int_0^1 dx\, J_0(2 \sqrt{\kappa t x}) \right\vert^2 = \frac{(J_1(2 \sqrt{\kappa t}))^2}{\kappa t} \, .
\end{equation}
Note that this result is actually independent of the precise density distribution as long as $k \sigma \gg 1$ and we have only chosen a uniform distribution to simplify the calculations. Consequently, in the limit $N \to \infty$ and $k \sigma \gg 1$, the dynamics of the bright state exactly reduces to the chiral case given by Eq.~(\ref{eq:pw_asymptotic}). The same result has already been found in \cite{Svidzinsky2010,Rohlsberger2013}, where the authors studied a similar system in three dimensions treating the atoms as point-like emitters and neglecting any polarization effects by taking only the scalar photon propagator. Further, their decay rate is increased by a factor of $2$ as they consider an initial excitation of forward- and backward-propagating modes.

The second regime of interested is obtained, if we assume, that the width of the distribution of the positions  is much smaller than the wavelength, i.e.,  $k \sigma \ll 1$. Then, all atoms are confined within one wavelength. In this limit, also $k \vert x_j - x_l \vert \ll 1$ for all $j$ and $l$. Thus, we might expand the sine and cosine term in Eqs.~(\ref{eq:exchange_1D_bidirectional}) and (\ref{eq:decay_1D_bidirectional}) leading to $J_{jl} \approx 0$ and $\Gamma_{jl} \approx 2\gamma$, respectively. Clearly, there is no coupling to the dark states while the bright state decays exponentially with a collectively enhanced decay rate $2 N \gamma$. The factor of $2$ appears because of the bidirectional coupling to the forward and backward-propagating modes. This limit corresponds to the situation Dicke considered originally where the particles are close to each other and single-photon superradiance is restored.
This result can also be derived analytically noting that in the limit $k \sigma \ll 1$, the effective Schr\"odinger equation (\ref{eq:effective_SE}) reduces to
\begin{equation}
    \partial_t \psi(x,t) = - \kappa \int dy \, n(y) \psi(y,t) \, ,
\end{equation}
with the same initial condition. As $\psi(x,t)$ varies only slowly within the range of $\sigma$, the differential equation is solved by the function $\psi(x,t) = \psi(x,0) e^{- \kappa t}$. Then, the population of the bright state is given by
\begin{equation}
    P_W(t) = \left\vert \frac{1}{N}\int dx \, n(x) e^{- \kappa t}\right\vert^2 = e^{- 2 \kappa t} \, ,
\end{equation}
with the collectively enhanced decay rate $\kappa = N \gamma$ \, .

\section{Conclusion}
\label{sec:conclusion}

In this paper, we have studied the dynamics of a single collective excitation of  $N$ two-level atoms influenced by photon-mediated coherent interactions. While one expects a collectively enhanced spontaneous decay if all the atoms are close together as predicted by Dicke in his seminal work \cite{Dicke1954}, we demonstrate within an analytical approach for a one-dimensional waveguide that the general long-time behavior is significantly modified due to the coherent exchange of virtual photons.  Especially, for large numbers of particles this exchange gives rise to a characteristic algebraic behavior $\sim 1/(\kappa t)^{3/2}$. The slowed down decay can be explained by the additional coupling of the superradiant bright state to subradiant states with different, but slower, decay rates. While this result is rigorous for a chiral waveguide, we demonstrate that this behavior also emerges in a bidirectional waveguide if the atoms are randomly distributed on a length scale larger than the optical wavelength of the transition. This opens up the possibility to study the peculiar influence of the virtual exchange of photons in systems without requiring a strictly chiral coupling which can be implemented much more easily. Examples of potential applications include, but are not limited to, atoms coupled to optical nanofibers or waveguides \cite{Petersen2014, Vetsch2010,Hood2016}, quantum dots coupled to photonic crystal waveguides or nanostructures \cite{Lodahl2004, Lodahl2015}, vacancy centers in crystals \cite{Sipahigil2016}, superconducting qubits \cite{VanLoo2013, Mirhosseini2019, Kannan2020} and also molecular chains \cite{Luo2019}.
Even though we focused our analysis on a purely one-dimensional system, we expect similar behavior to also appear in three-dimensional setups in free space where the atoms are coupled to a single focused light mode. Our observations are thus relevant for a broad range of systems with collective excitations as for example quantum memories. In particular, it is of fundamental importance for understanding Rydberg superatoms in free space which have recently attracted a lot of experimental attention \cite{Paris-Mandoki2017}, and we expect that the influence of this coherent exchange interaction is also relevant for the recent experimental observation of an oscillatory behavior of the decay rate of such Rydberg superatoms \cite{Bettles2018, Stiesdal2020}.

\section{Acknowledgements}
This work is supported by the European
Union’s Horizon 2020 program under the ERC consolidator
grants SIRPOL (grant N. 681208) and RYD-QNLO (grant N.
771417), the ErBeStA project (No. 800942), grant agreement
No. 845218 (Marie Skłodowska-Curie Individual Fellowship
to H.B.), and the Deutsche Forschungsgemeinschaft (DFG)
under SPP 1929 GiRyd project BU 2247/4-1 and the research unit FOR 2247.


\appendix

\section{Master equation and photon propagator}

\subsection{Derivation of the master equation}
\label{app:master_equation}

Here, we give some additional information on how to arrive at Eq.~(\ref{eq:master_equation}) in the main text and will explicitly calculate the propagator $G$ for the case of a one-dimensional waveguide with both chiral and bidirectional coupling.

The electric field operator at any given point in space and time consists of the incoming field $\mathcal{E}_0(r, t)$ and the field due to the scattering off the emitters (i.e., dipoles). Within the narrow-bandwidth approximation and by neglecting retardation it is
\begin{equation}
    \mathcal{E}^-(r, t) = \mathcal{E}_0(r, t) + \sqrt{\gamma} \sum_{j=1}^N G(r, r_j, \omega_0) \sigma_j^-(t) \, ,
\end{equation}
where we assume the two-level emitters with transition frequency $\omega_0$ to sit at positions $r_j$ and the coupling between them and the light field is $\sqrt{\gamma}$. We now determine the master equation for the emitter subsystem in terms of the propagator $G$.

In the dipole and rotating-wave approximation, the interaction between the light field and the emitters is given by
\begin{equation}
    H_\text{int} = -\hbar \sqrt{\gamma} \sum_{j=1}^N \mathcal{E}^+(r_j) \sigma_j^- + \sigma_j^+ \mathcal{E}^-(r_j) \, .
\end{equation}
For an arbitrary operator $O$ that acts only on the subsystem of the emitters, we get the Heisenberg equation of motion
\begin{widetext}
\begin{align}
    \partial_t O = \frac{i}{\hbar}[H_\text{int}, O] &= -i \sqrt{\gamma} \sum_{j=1}^N \mathcal{E}^+(r_j) [\sigma_j^-, O] + [\sigma_j^+,O] \mathcal{E}^-(r_j) \nonumber \\
    &= -i \sqrt{\gamma} \sum_{j=1}^N \mathcal{E}_0^+(r_j) [\sigma_j^-, O] + [\sigma_j^+, O] \mathcal{E}_0^-(r_j) + i \gamma \sum_{j,l=1}^N G^*(r_j, r_l, \omega_0) \sigma_l^+ [\sigma_j^-, O] +  G(r_j, r_l, \omega_0)  [\sigma_j^+, O]\sigma_l^- \, .
\end{align}
Note that $\mathcal{E}_0$ gives the free evolution of the incoming field and does not depend on any emitter operators. It can therefore be added to $H_0$ as a classical driving field by using the Mollow transformation \cite{Mollow1975}. Consequently, we will neglect this contribution in the following.

Using that $\partial_t \langle O \rangle = \text{tr}(O \partial_t \rho)$, we can derive the equation of motion for the reduced density matrix of the emitters by
\begin{align}
    \partial_t \langle O(t) \rangle &= \text{tr}(O \partial_t \rho(t)) \nonumber \\
    &= -i \gamma \sum_{j=1}^N G^*(r_j, r_l, \omega_0) \text{tr}( \sigma_l^+ [ \sigma_j^-, O] \rho(t) ) + G(r_j, r_l, \omega_0) \text{tr}( [\sigma_j^+, O] \sigma_l^- \rho(t) ) \nonumber \\
    & = -i \gamma \sum_{j,l=1}^N G^*(r_j, r_l, \omega_0) \text{tr}( \sigma_l^+ \sigma_j^- O \rho(t) - \sigma_l^+ O \sigma_j^- \rho(t)) + G(r_j, r_l, \omega_0) \text{tr}( \sigma^+_j O \sigma_l^- \rho(t) - O \sigma_j^+ \sigma_l^- \rho(t) ) \nonumber \\
    & = -i \gamma \sum_{j,l=1}^N G^*(r_j, r_l, \omega_0) \text{tr}(O(\rho(t) \sigma_l^+ \sigma_j^- - \sigma_j^- \rho(t) \sigma_l^+)) + G(r_j, r_l, \omega_0) \text{tr}(O(\sigma_l^- \rho(t) \sigma_j^+ - \sigma_j^+ \sigma_l^- \rho(t))) \, .
\end{align}
Since the operator $O$ is arbitrary, we can infer the equation of motion for the density matrix $\rho$,
\begin{align}
    \partial_t \rho(t) &= -i \gamma \sum_{j,l=1}^N G^*(r_j, r_l, \omega_0) ( \rho(t) \sigma_l^+ \sigma_j^- - \sigma_j^- \rho(t) \sigma_l^+) + G(r_j, r_l, \omega_0) (\sigma_l^- \rho(t) \sigma_j^+ - \sigma_j^+ \sigma_l^- \rho(t)) \nonumber \\
    & = -i \gamma \sum_{j,l=1}^N \left\lbrace (G(r_j,r_l, \omega_0) - G^*(r_l, r_j, \omega_0)) \sigma_l^- \rho(t) \sigma_j^+  - \frac{1}{2} ( G(r_j, r_l, \omega_0) + G^*(r_l, r_j, \omega_0)) [\sigma_j^+ \sigma_l^-, \rho(t)] \right. \nonumber \\
    & \qquad \left.- \frac{1}{2}(G(r_j, r_l, \omega_0) - G^*(r_l, r_j, \omega_0)) \{ \sigma_j^+ \sigma_l^-, \rho(t)\} \right\rbrace \nonumber \\
    & = -\frac{i}{\hbar} \left[  \hbar \sum_{j,l=1}^N J_{jl} \sigma_j^+ \sigma_l^-, \rho \right] + \sum_{j,l=1}^N \Gamma_{jl} \left( \sigma_l^- \rho(t) \sigma_j^+ - \frac{1}{2} \{\sigma_j^+ \sigma_l, \rho(t) \}\right) \, ,
\end{align}
with the exchange interaction terms and decay rates
\begin{align}
    J_{jl} &= -\gamma \frac{G^*(r_l, r_j, \omega_0) + G(r_j, r_l, \omega_0)}{2} \, , \\
    \Gamma_{jl} &= i \gamma (G^*(r_l, r_j, \omega_0) - G(r_j, r_l, \omega_0)) \, .
\end{align}

\subsection{Photon propagator of a one-dimensional waveguide}
\label{app:photon_propagator}

\subsubsection{Chiral waveguide}

In order to derive the propagator for the one-dimensional chiral waveguide, we start with the Hamiltonian in the dipole and rotating-wave approximation and in the rotating frame of the atoms
\begin{equation}
    H = \hbar \int_{k - q_c}^{k+q_c} \frac{dq}{2\pi} \omega_q a_q^\dagger a_q  - \hbar \sqrt{\gamma} \sum_{j=1}^N \mathcal{E}^+(x_j) \sigma_j^- + \sigma_j^+ \mathcal{E}^-(x_j) \, ,
\end{equation}
where $a_q^{(\dagger)}$ annihilate (create) photons with momentum $q$ having a dispersion relation $\omega_q = c q - \omega_0$, with the resonance frequency of the atomic transition, $\omega_0$. Note that we only integrate over the relevant modes of the waveguide, which are centered around $|k| = \omega_0/c$, and, since the waveguide is chiral, we only consider forward propagating modes with positive momenta. The last term describes the interaction of the waveguide photons with the emitters with an effective mode coupling $\sqrt{\gamma}$. The electric field operator reads
\begin{equation}
    \mathcal{E}^-(x) = i \sqrt{c} \int_{k-q_c}^{k+q_c} \frac{dq}{2\pi} a_q e^{i q x} \, ,
    \label{eqsupp:electric_field}
\end{equation}

The time evolution of the electric field can be calculated using the Heisenberg equation of motion
\begin{equation}
    \dot{a}_q(t) = \frac{i}{\hbar} [H, a_q] = - i \omega_q a_q(t) +\sqrt{\gamma \, c} \sum_{j=1}^N e^{- i q x_j} \sigma_j^-(t) \, .
\end{equation}
This differential equation can be formally solved by integration which leads to
\begin{equation}
    a_q(t) = a_q(0) e^{- i \omega_q t} + \sqrt{\gamma \, c} \sum_{j=1}^N e^{- i q x_j} \int_0^t ds \, e^{-i \omega_q(t - s)} \sigma_j^-(s) \, .
\end{equation}
Plugging this expression back into the mode expansion of the electric field, Eq.~(\ref{eqsupp:electric_field}), gives
\begin{align}
    \mathcal{E}^-(x,t) &= i \sqrt{c} \int_{k - q_c}^{k+q_c} \frac{dq}{2\pi} \eta_q a_q(0) e^{- i \omega_q t + i q x} + i c \sqrt{\gamma} \sum_{j=1}^N \int_{k - q_c}^{k+q_c} \frac{dq}{2\pi}  e^{i q (x-x_j)} \int_0^t ds \, e^{- i \omega_q(t-s)} \sigma_j^-(s) \nonumber \\
    & = \mathcal{E}^-_0(x,t) + i c \sqrt{\gamma} \sum_{j=1}^N \int_{k - q_c}^{k+q_c} \frac{dq}{2\pi}  e^{i q (x-x_j)} \int_0^t ds \, e^{- i \omega_q(t-s)} \sigma_j^-(s)\, ,
    \label{eqsupp:electric_field_time}
\end{align}
where $\mathcal{E}^-_0$ describes the non-interacting component of the electric field.

In order to further simplify the expression for the electric field, we change from an integration over the momentum to an integration over the frequency, where $\omega = c q$ and $\omega_c = c q_c$, such that
\begin{align}
    i c \sqrt{\gamma} \sum_{j=1}^N \int_{k - q_c}^{k+q_c} \frac{dq}{2\pi}  e^{i q (x-x_j)} \int_0^t ds \, e^{- i \omega_q(t-s)} \sigma_j^-(s) & = i \sqrt{\gamma} \sum_{j=1}^N \int_{\omega_0 - \omega_c}^{\omega_0 + \omega_c} \frac{d \omega}{2\pi} e^{i \omega(x- x_j)/c} \int_0^t ds \, e^{-i (\omega - \omega_0)(t-s)} \sigma_j^-(s) \nonumber \\
    & = i \sqrt{\gamma} \sum_{j=1}^N e^{i \omega_0 (x-x_j)/c} \int_0^t ds \, \frac{\sin(\omega_c(t-s-(x-x_j)/c))}{\pi(t-s-(x-x_j)/c)} \sigma_j^-(s) \, .
    \label{eq:sin_integral}
\end{align}
The last expression can be simplified by assuming that the atomic operators $\sigma_j^-$ only slowly vary on a time scale $N \gamma$ with $N \gamma \ll \omega_c \ll \omega_0$. The integral over the time then only contributes significantly when $s = t - (x-x_j)/c$ as long as $x \geq x_j$ and we can approximate the time integral in the last expression in Eq. (\ref{eq:sin_integral}) as
$ \theta(x - x_j)\sigma_j^-(t - (x-x_j)/c) \, ,$ where $\theta(x)$ is the Heavside function with $\theta(x) = 1$ if $x > 0$, $\theta(x) = 0$ if $x<0$ and $\theta(0) = 1/2$. The above approximation is known as narrow-bandwidth approximation and is closely connected to the Markov approximation \cite{Gardiner2004}. The electric field can then be written as
\begin{align}
    \mathcal{E}^-(x,t) & = \mathcal{E}^-_0(x,t) \nonumber \\
    &\, + i \sqrt{\gamma} \sum_{j=1}^N \theta(x-x_j) e^{i k(x-x_j)} \sigma_j^-(t - (x-x_j)/c) \, .
\end{align}
Note that this expression still includes retardation effects. However, these can be neglected if $N\gamma \ll c / \vert x - x_j \vert $, i.e. if the time scale for the propagation of a photon through the waveguide is much smaller than the time scale on which the atomic operators evolve. Then, we can approximate $\sigma_j^-(t - (x-x_j)/c) \approx \sigma_j^-(t)$. Finally, the expression for the electric field reads
\begin{align}
    \mathcal{E}^-(x,t) = \mathcal{E}^-_0(x,t) + i \sqrt{\gamma} \sum_{j=1}^N \theta(x-x_j) e^{i k(x-x_j)} \sigma_j^-(t)
\end{align}
and we can identify the propagator as
\begin{align}
    G(x,x_j, \omega_0) = i \theta(x-x_j) e^{i \omega_0 (x-x_j)/c} \, .
\end{align}

\subsubsection{Bidirectional waveguide}

For a bidirectional waveguide, the calculation is very similar to the case discussed above. In contrast to the chiral setup, the Hamiltonian describing the waveguide photons now reads
\begin{align}
    H_0 = \hbar \int_{k - q_c}^{k + q_c} \frac{dq}{2\pi} \omega_q a_q^\dagger a_q + \hbar \int_{-k - q_c}^{-k + q_c} \frac{dq}{2\pi} \omega_q a_q^\dagger a_q
\end{align}
as we are coupling to both forward- and backward-propagating modes with positive and negative momenta, respectively. The electric field operator analogously is
\begin{equation}
    \mathcal{E}^-(x) = i \sqrt{c} \left( \int_{k - q_c}^{k + q_c} \frac{dq}{2\pi} + \int_{-k - q_c}^{-k + q_c} \frac{dq}{2\pi}\right) e^{i q x}a_q \, .
    \label{eqsupp:E-field_bidirectional}
\end{equation}

Similar to above, we can derive the Heisenberg equation of motion for the photonic operator $a_q$, formally integrate it and plug it into the expression for the electric field, Eq. (\ref{eqsupp:E-field_bidirectional}). Changing from an integration over momenta to an integration over frequencies, we get for the interaction part

\begin{align}
    i c \sqrt{\gamma} \sum_{j=1}^N \left(\int_{k - q_c}^{k+q_c} + \int_{-k - q_c}^{-k+q_c}\right) \frac{dq}{2\pi}  e^{i q (x-x_j)} \int_0^t ds \, e^{- i \omega_q(t-s)} \sigma_j^-(s) & = i \sqrt{\gamma} \sum_{j=1}^N \sum_{\lambda = \pm} \int_{\omega_0 - \omega_c}^{\omega_0 + \omega_c} \frac{d\omega}{2\pi} e^{i \lambda \omega(x-x_j)/c} \nonumber \\
    & \quad \times \int_0^t ds \, e^{-i (\omega - \omega_0)(t-s)} \sigma_j^-(s) \, .
\end{align}
The only difference in the bidirectional case now is that we have in addition to sum over two different modes $\lambda = \pm$. Along the same lines as in the chiral case, we get
\begin{align}
    i \sqrt{\gamma} \sum_{j=1}^N \sum_{\lambda = \pm} \int_{\omega_0 - \omega_c}^{\omega_0 + \omega_c} \frac{d\omega}{2\pi} e^{i \lambda \omega(x-x_j)/c}  \int_0^t ds \, e^{-i (\omega - \omega_0)(t-s)} \sigma_j^-(s) &\approx i \sqrt{\gamma} \sum_{j=1}^N \sum_{\lambda = \pm} \theta(\lambda(x - x_j)) e^{i \lambda k (x-x_j)} \sigma_j^-(t) \nonumber \\
    & = i \sum_{j=1}^N e^{i k \vert x - x_j \vert} \sigma_j^-(t) \, .
\end{align}

Consequently, the propagator for the bidirectional waveguide is
\begin{equation}
    G(x, x_j, \omega_0) = i e^{ i \omega_0 \vert x - x_j \vert /c} \, .
\end{equation}

\end{widetext}

\section{Analytical solution for $P_W(t)$ for $N$ atoms}
\label{app:analytical}

In this section, we present two possible ways to derive Eq.(\ref{eq:brightStateDecay}) from the main text. First, we use the method of the Bethe Ansatz used in~\cite{Yudson1985}. As an alternative approach, we present the derivation using an effective Hamiltonian.

\subsection{Solution using the Bethe Ansatz}
The decay profile of the bright state, Eq.~(\ref{eq:brightStateDecay}) in the main text, can be derived from the
microscopic theory described by the Hamiltonian~(\ref{eq:hamiltonian}) in the main text. This
approach gives an alternative point of view and validates any approximation (e.g., the
Wigner-Weisskopf approximation) in the derivation of the electrical field
propagator, Eq.~(\ref{eq:electric_field}) in the main text. We assume a linear dispersion relation for the
photons. This in turn, allows to solve the full Hamiltonian with the Bethe Ansatz, as
demonstrated in~\cite{Yudson1985}. The eigenstates with a single excitation are given by
\begin{widetext}
\begin{align}
	|\lambda\rangle &= \int \frac{dy}{\sqrt{2\pi}}
	\prod_{j=1}^N\frac{\lambda-i\gamma/2\operatorname{sgn}(y-x_j)}{\lambda+i\gamma/2}e^{i\lambda x}
		\left( b^\dagger(x) - \frac{\sqrt{\gamma}}{\lambda} \sum_j \delta(y-x_j)\sigma_j^+\right) |0\rangle.
	\label{eq:SingleExcitationState}
\end{align}
\end{widetext}
Here, $\lambda$ may be interpreted as the momentum of the excitation.

We determine the time evolution of the bright state by decomposing it in the basis of Bethe
states. As a first step, however, we start with the time evolution of a single excited atom $|\psi_j\rangle = \sigma_j^+|0\rangle$ and project it onto the excited state $|\psi_l\rangle$,
\begin{align}
	\notag
	\langle\psi_l|\psi_j(t)\rangle &= \int_{-\infty}^\infty d\lambda
		e^{-i\lambda t} \langle\psi_l|\lambda\rangle\langle\lambda|\psi_j\rangle \\
		&= -i\gamma^2 L_{l-j}^{(-1)}(\gamma t) e^{-\gamma t/2}
	\label{eq:DecayIthAtom}
\end{align}
The time evolution of the $|W\rangle$ may now be calculated by summing the
individual evolution of each atom in $|W\rangle$ and projecting back onto the atoms ($P_\mathrm{atoms}$):
\begin{align}
	\notag
	P_\mathrm{atoms}|W(t)\rangle
		&= P_\mathrm{atoms} \frac{1}{\sqrt{N}} \sum_{j=1}^N U(t) \sigma_j |0\rangle \\
	\notag
	&= \frac{i\gamma^2}{\sqrt{N}} \sum_{j=1}^N \left( \sum_{l\ge j} L_{l-j}^{(-1)}(\gamma t) \right)
		e^{-\gamma t/2} \sigma_j |0\rangle \\
	&= \frac{i\gamma^2}{\sqrt{N}} \sum_{j=1}^N L_{j-1}(\gamma t) e^{-\gamma t/2} \sigma_j |0\rangle.
	\label{eq:TimeEvoWState}
\end{align}
The result given by Eq.~(\ref{eq:brightStateDecay}) in the main text for the decay of the
$|W\rangle$ state readily follows
\begin{align}
	P_{W}(t) &=
	    \left[ \frac{1}{N} \sum_{j=1}^N L_{j-1}(\gamma t) \right]^2 e^{-\gamma t} \\
	&= \left[ \frac{1}{N} L_{N-1}^{(1)}(\gamma t) \right]^2 e^{-\gamma t}.
	\label{eq:PW}
\end{align}
Analogously, the probability to have any atom excited is the squared norm of the
$|W(t)\rangle$ state
\begin{align}
	\notag
	\langle P_\mathrm{atoms}(t)\rangle
	    &= \frac{1}{N} \sum_{j=1}^N L_{j-1}^2(\gamma t) e^{-\gamma t} \\
	&= \left[ L_{N-1}(\gamma t) L_N(\gamma t) - L_{N-1}^{(1)}(\gamma t) L_N^{(-1)}(\gamma t) \right]
		e^{-\gamma t}.
	\label{eq:PAtoms}
\end{align}

For large $N$ the Laguerre polynomials $L_N^{(\alpha)}(x)$ are well approximated by Bessel functions
\begin{equation}
	L_N^{(\alpha)}(x) \approx \sqrt{N^\alpha}\frac{J_\alpha(2\sqrt{N x})}{\sqrt{x^\alpha}} e^{x/2},
	\label{eq:ApproxLaguerre}
\end{equation}
which are, for large $x$, approximated by an algebraic decay, superimposed with a harmonic
oscillation
\begin{equation}
	J_\alpha(x) \approx \sqrt{\frac{2}{\pi x}} \cos\left(x - \frac{\alpha\pi}{2} - \frac{\pi}{4} \right).
	\label{eqApproxBessel}
\end{equation}
Hence, for many atoms $N \gg 1$ and for times $\gamma t > 1$ we find the asymptotic
expressions
\begin{equation}
	P_{W}(t) \approx \frac{1}{\pi\sqrt{(\gamma N t)^3}}
		\cos^2\left( 2\sqrt{\gamma N t} - \frac{3\pi}{4} \right)
	\label{eq:PWAprox}
\end{equation}
and
\begin{equation}
	\langle P_{\mathrm{atoms}}(t)\rangle \approx \frac{1}{\pi\sqrt{\gamma N t}}.
	\label{eq:PTotalAprox}
\end{equation}

\subsection{Solution using the effective Hamiltonian}

In this section, we present the derivation of Eq.(\ref{eq:brightStateDecay}) from the main text using the effective (non-Hermitian) Hamiltonian
\begin{equation}
H_\text{eff} = \frac{\hbar \gamma}{2 i} \sum_{j,l} \left(\text{sign}(x_j-x_l) +1 \right)e^{i k (x_j - x_l)} \sigma^+_j \sigma^-_l \, .
\end{equation}
Even though the emitter system is described by a master equation, in the absence of driving and assuming the system is initially prepared in the bright state $\ket{W} = \frac{1}{\sqrt{N}} \sum_j e^{i k x_j} \sigma_j^+ \ket{G}$, it is possible to describe the time evolution of $\ket{W}$ with the effective Hamiltonian above.

In order to simplify the calculations, we absorb all phases into the operators, that is $\sigma_j^+  \to e^{- i k x_j} \sigma_j^+$ and similarly for $\sigma_j^-$. The effective Hamiltonian can then be written as
\begin{equation}
H_\text{eff} = \frac{\hbar \gamma}{2 i} \sum_{j,l} \left(\text{sign}(x_j-x_l) +1 \right)\sigma^+_j \sigma^-_l \, .
\label{eq:Heff}
\end{equation}
Note that this transformation is not useful in the case of a bidirectional system and reflects the fact that for a chiral system only the order of the emitters is important but not their relative distance. In the following, we will assume that $x_j < x_l$ if $j < l$.

In the basis $\{ \ket{j} = \sigma_j^+ \ket{G}, \, j = 1, \ldots, N \}$, we can represent the Hamiltonian Eq.~(\ref{eq:Heff}) as the sum of the $N \times N$ identity matrix $I$ and a nilpotent matrix $M_N$ for which $(M_N)^n = 0 , \, n\geq N$:
\begin{equation}
H_\text{eff} = -\frac{i \hbar \gamma}{2} (I + 2 M_N) \,
\end{equation}
with
\begin{equation}
M_N = \begin{pmatrix}
0 & 0 & \cdots & 0 \\
1 & 0 & \cdots &0 \\
\vdots & \vdots & \ddots & \vdots \\
1 & 1 & \cdots & 0
\end{pmatrix} \, .
\end{equation}

The time evolution of the bright state is then given by
\begin{align}
\ket{\psi(t)} &= e^{- i H_\text{eff} t/ \hbar} \ket{W} = e^{- \frac{\gamma t}{2} I} \sum_{n= 0}^{N-1} \frac{(-\gamma t)^n}{n!} (M_N)^n \ket{W} \, .
\end{align}

The probability to remain in the bright state as a function of time can be written as
\begin{equation}
P_W(t) =e^{-\gamma t} \left\vert  \sum_{n=0}^{N-1} \frac{(-\gamma t)^n}{n!} \bra{W} (M_N)^n \ket{W} \right\vert^2\, .
\end{equation}

In the basis given above, $\ket{W}$ is represented by the vector
\begin{equation}
\ket{W} = \frac{1}{\sqrt{N}} \begin{pmatrix}
1 \\
\vdots \\
1
\end{pmatrix} \,
\end{equation}
such that the matrix element $\bra{W} (M_N)^n \ket{W}$ can be calculated as
\begin{align}
\bra{W} (M_N)^n \ket{W} &= \sum_{j_1 < j_2 < \cdots < j_{n+1}} \frac{1}{N} \nonumber \\
& = \frac{1}{N} \frac{N(N-1) \cdots (N-n)}{(n+1)!} \nonumber \\
& = \frac{1}{N} \begin{pmatrix}
N \\
n+1
\end{pmatrix} = \frac{1}{N} \begin{pmatrix}
N \\
N - (n+1)
\end{pmatrix} \, .
\end{align}

Finally, the time evolution of the occupation of the bright state reads
\begin{align}
P_W(t) & = \vert \bra{W} e^{- i H_\text{eff} t / \hbar} \ket{W} \vert^2 \nonumber \\
& =  \left\vert \sum_{n=0}^{N-1} \frac{(-\gamma t)^n}{n!} \frac{1}{N} \begin{pmatrix}
N \\
N - (n+1)
\end{pmatrix}  \right\vert^2 e^{-\gamma t} \nonumber \\
& = \left(\frac{1}{N} L_{N-1}^{(1)} ( \gamma t) \right)^2 e^{-\gamma t}
\end{align}
where $L_n^{(\alpha)}(x)$ is the generalized Laguerre polynomial. This is exactly the same result as obtained using the Bethe ansatz above.

\section{Continuum limit in the bidirectional case}
\label{app:continuum}

Here, we show that the time evolution of the bright state in the bidirectional waveguide reduces to the time evolution in the chiral case in the limit where $N \to \infty$ and $k\sigma \to \infty$. In contrast to the numerical calculations mentioned in the main text, we assume the atoms to be uniformly distributed in an interval $[0,\sigma]$ along the waveguide such that the analytical calculations simplify. The final result, however,  does not depend on the details of the distribution as long as $k \sigma \gg 1$.

In the limit $N \to \infty$ and $\sigma$ finite, we can go over to the continuum limit by keeping $\kappa = N \gamma$ fixed. The effective Hamiltonian in this case reads
\begin{equation}
H = -i \gamma \int dx\, dy \, \exp(i k \vert x - y \vert) \Psi^\dagger(x) \Psi(y)
\end{equation}
with the field creation and annihilation operators $\Psi^\dagger(x)$ and $\Psi(x)$, respectively. The have the commutation relations $[\Psi(x), \Psi^\dagger(y)] = \delta(x-y)$. The initial bright state is given by
\begin{equation}
\ket{W} = \frac{1}{\sqrt{N}} \int dx \, e^{i k x} \Psi^\dagger(x) \ket{G} \, .
\end{equation}
In order to calculate the time evolution for the state
\begin{equation}
\ket{\psi(t)} = \frac{1}{\sqrt{N}} \int dx \, \psi(x,t) \Psi^\dagger(x) \ket{G}\, ,
\end{equation}
we have to solve the effective Schr\"odinger equation
\begin{equation}
i \partial_t \psi(x,t) = - i \frac{\kappa}{\sigma} \int_0^\sigma dy \, \exp(i k \vert x - y \vert) \psi(y,t) \, .
\end{equation}
In the following, we rescale all lengths by $\sigma$ and introduce the dimensionless quantity $q = k \sigma$. Further, we rescale all times by the collective rate $\kappa$. Then, the dimensionless Schr\"odinger equation reads
\begin{equation}
\partial_t \psi(x,t) = - \int_0^1 dy \, \exp(i q \vert x - y \vert) \psi(y,t) \,
\end{equation}
with the initial condition $\psi(x,0) = e^{ikx}$.

In order to solve this differential equation, we first apply a Laplace transform from the variable $t$ to the variable $s$,
\begin{widetext}
\begin{align}
s \, \hat{\psi}(x,s) - \psi(x,0) &= - \int_0^1 dy \, e^{i q \vert x - y \vert} \hat{\psi}(x,s) \,  \nonumber \\
&= - \int_0^x dy \, e^{i q (x-y)} \hat{\psi}(x,s) - \int_x^1 dy \, e^{- i q(x-y)} \hat{\psi}(x,s) \, \nonumber \\
&= - \int_0^x dy \, e^{i q (x-y)} \hat{\psi}(y,s) + \int_0^x dy \, e^{- i q(x-y)} \hat{\psi}(y,s) - \int_0^1 dy \, e^{-i q (x-y)} \hat{\psi}(y,s) \, .
\end{align}
\end{widetext}

As we want to get rid of fast oscillating terms in the end, we make the ansatz $\hat{\psi}(x,s) = e^{i q x} \hat{\phi}(x,s)$, where $\hat{\phi}(x,s)$ is assumed to be a slowly varying function of $x$. It then follows
\begin{align}
s \, \hat{\phi}(x,s) - 1 &= - \int_0^x dy \, \hat{\phi}(y,s) + \int_0^x dy \, e^{-2i q(x-y)} \hat{\phi}(y,s) \nonumber \\
& \quad -e^{-2 i q x} \int_0^1 dy \, e^{2i q y} \hat{\phi}(y,s) \, .
\end{align}
The last integral in this expression vanishes in the limit $q\to \infty$ and we can drop it in the following. Next, we apply a Laplace transform from the variable $x$ to $u$ which leads to
\begin{align}
s \hat{\hat{\phi}} (u,s) -\frac{1}{u} = - \frac{\hat{\hat{\phi}}(u,s)}{u} + \frac{\hat{\hat{\phi}}(u,s)}{u+2 i q} \, ,
\end{align}
where we have made use of the convolution theorem for the Laplace transform. The integral equation is then reduced to an algebraic one whose solution reads
\begin{equation}
\hat{\hat{\phi}}(u,s) = \frac{u+2iq}{u(u+2iq)(s + \frac{1}{u} - 1)} \, .
\end{equation}
Now we can take the limit $q \to \infty$ and are left with
\begin{equation}
\hat{\hat{\phi}}(u,s) \approx \frac{1}{s u + 1} \, .
\end{equation}
The inverse Laplace transform of this expression back to the variables $x$ and $t$ is given by
\begin{equation}
\phi(x,t) = J_0(2 \sqrt{x t}) \,
\end{equation}
with the Bessel function of the first kind $J_0(x)$. Thus, the full solution for the wavefunction reads
\begin{equation}
\psi(x,t) = e^{i q x} J_0(2 \sqrt{ x t}) \, .
\end{equation}
The time evolution of the bright state is thus given by
\begin{align}
P_W(t) & = \left\vert \int_0^1 dx \, J_0(2 \sqrt{ x t}) \right\vert^2 \nonumber \\
& = \left\vert \frac{J_1(2 \sqrt{t})}{\sqrt{t}} \right\vert^2 \nonumber \\
& = \frac{(J_1(2 \sqrt{\kappa t}))^2}{\kappa t} \, ,
\label{eq:supp_pw_continuum}
\end{align}
where $J_1(x)$ is the Bessel function of the first kind and we reintroduced dimensioned variables. This is the same result as in the chiral case in the limit $N \to \infty$.

The probability of finding an excitation in the system at time $t$ is given by
\begin{align}
P(t) & = \int_0^1 dx \, (J_0(2 \sqrt{xt}))^2 \nonumber \\
& = J_0(2 \sqrt{ \kappa t})^2 + J_1(2 \sqrt{\kappa t})^2 \, ,
\end{align}
where again we have reintroduced dimensioned variables in the last line.

\bibliographystyle{apsrev4-1}
\bibliography{main_final}

\begin{thebibliography}{63}%
\makeatletter
\providecommand \@ifxundefined [1]{%
 \@ifx{#1\undefined}
}%
\providecommand \@ifnum [1]{%
 \ifnum #1\expandafter \@firstoftwo
 \else \expandafter \@secondoftwo
 \fi
}%
\providecommand \@ifx [1]{%
 \ifx #1\expandafter \@firstoftwo
 \else \expandafter \@secondoftwo
 \fi
}%
\providecommand \natexlab [1]{#1}%
\providecommand \enquote  [1]{``#1''}%
\providecommand \bibnamefont  [1]{#1}%
\providecommand \bibfnamefont [1]{#1}%
\providecommand \citenamefont [1]{#1}%
\providecommand \href@noop [0]{\@secondoftwo}%
\providecommand \href [0]{\begingroup \@sanitize@url \@href}%
\providecommand \@href[1]{\@@startlink{#1}\@@href}%
\providecommand \@@href[1]{\endgroup#1\@@endlink}%
\providecommand \@sanitize@url [0]{\catcode `\\12\catcode `\$12\catcode
  `\&12\catcode `\#12\catcode `\^12\catcode `\_12\catcode `\%12\relax}%
\providecommand \@@startlink[1]{}%
\providecommand \@@endlink[0]{}%
\providecommand \url  [0]{\begingroup\@sanitize@url \@url }%
\providecommand \@url [1]{\endgroup\@href {#1}{\urlprefix }}%
\providecommand \urlprefix  [0]{URL }%
\providecommand \Eprint [0]{\href }%
\providecommand \doibase [0]{http://dx.doi.org/}%
\providecommand \selectlanguage [0]{\@gobble}%
\providecommand \bibinfo  [0]{\@secondoftwo}%
\providecommand \bibfield  [0]{\@secondoftwo}%
\providecommand \translation [1]{[#1]}%
\providecommand \BibitemOpen [0]{}%
\providecommand \bibitemStop [0]{}%
\providecommand \bibitemNoStop [0]{.\EOS\space}%
\providecommand \EOS [0]{\spacefactor3000\relax}%
\providecommand \BibitemShut  [1]{\csname bibitem#1\endcsname}%
\let\auto@bib@innerbib\@empty
\bibitem [{\citenamefont {Guerin}\ \emph {et~al.}(2017)\citenamefont {Guerin},
  \citenamefont {Rouabah},\ and\ \citenamefont {Kaiser}}]{Guerin2016a}%
  \BibitemOpen
  \bibfield  {author} {\bibinfo {author} {\bibfnamefont {W.}~\bibnamefont
  {Guerin}}, \bibinfo {author} {\bibfnamefont {M.}~\bibnamefont {Rouabah}}, \
  and\ \bibinfo {author} {\bibfnamefont {R.}~\bibnamefont {Kaiser}},\ }\href
  {\doibase 10.1080/09500340.2016.1215564} {\bibfield  {journal} {\bibinfo
  {journal} {Journal of Modern Optics}\ }\textbf {\bibinfo {volume} {64}},\
  \bibinfo {pages} {895} (\bibinfo {year} {2017})},\ \Eprint
  {http://arxiv.org/abs/1605.02439} {arXiv:1605.02439} \BibitemShut {NoStop}%
\bibitem [{\citenamefont {Dicke}(1954)}]{Dicke1954}%
  \BibitemOpen
  \bibfield  {author} {\bibinfo {author} {\bibfnamefont {R.~H.}\ \bibnamefont
  {Dicke}},\ }\href {\doibase 10.1103/PhysRev.93.99} {\bibfield  {journal}
  {\bibinfo  {journal} {Physical Review}\ }\textbf {\bibinfo {volume} {93}},\
  \bibinfo {pages} {99} (\bibinfo {year} {1954})}\BibitemShut {NoStop}%
\bibitem [{\citenamefont {Gross}\ and\ \citenamefont
  {Haroche}(1982)}]{Gross1982}%
  \BibitemOpen
  \bibfield  {author} {\bibinfo {author} {\bibfnamefont {M.}~\bibnamefont
  {Gross}}\ and\ \bibinfo {author} {\bibfnamefont {S.}~\bibnamefont
  {Haroche}},\ }\href {\doibase 10.1016/0370-1573(82)90102-8} {\bibfield
  {journal} {\bibinfo  {journal} {Physics Reports}\ }\textbf {\bibinfo {volume}
  {93}},\ \bibinfo {pages} {301} (\bibinfo {year} {1982})}\BibitemShut
  {NoStop}%
\bibitem [{\citenamefont {R{\"{o}}hlsberger}\ \emph {et~al.}(2010)\citenamefont
  {R{\"{o}}hlsberger}, \citenamefont {Schlage}, \citenamefont {Sahoo},
  \citenamefont {Couet},\ and\ \citenamefont {Ruffer}}]{Rohlsberger2010}%
  \BibitemOpen
  \bibfield  {author} {\bibinfo {author} {\bibfnamefont {R.}~\bibnamefont
  {R{\"{o}}hlsberger}}, \bibinfo {author} {\bibfnamefont {K.}~\bibnamefont
  {Schlage}}, \bibinfo {author} {\bibfnamefont {B.}~\bibnamefont {Sahoo}},
  \bibinfo {author} {\bibfnamefont {S.}~\bibnamefont {Couet}}, \ and\ \bibinfo
  {author} {\bibfnamefont {R.}~\bibnamefont {Ruffer}},\ }\href {\doibase
  10.1126/science.1187770} {\bibfield  {journal} {\bibinfo  {journal}
  {Science}\ }\textbf {\bibinfo {volume} {328}},\ \bibinfo {pages} {1248}
  (\bibinfo {year} {2010})}\BibitemShut {NoStop}%
\bibitem [{\citenamefont {Pellegrino}\ \emph {et~al.}(2014)\citenamefont
  {Pellegrino}, \citenamefont {Bourgain}, \citenamefont {Jennewein},
  \citenamefont {Sortais}, \citenamefont {Browaeys}, \citenamefont {Jenkins},\
  and\ \citenamefont {Ruostekoski}}]{Pellegrino2014}%
  \BibitemOpen
  \bibfield  {author} {\bibinfo {author} {\bibfnamefont {J.}~\bibnamefont
  {Pellegrino}}, \bibinfo {author} {\bibfnamefont {R.}~\bibnamefont
  {Bourgain}}, \bibinfo {author} {\bibfnamefont {S.}~\bibnamefont {Jennewein}},
  \bibinfo {author} {\bibfnamefont {Y.~R.~P.}\ \bibnamefont {Sortais}},
  \bibinfo {author} {\bibfnamefont {A.}~\bibnamefont {Browaeys}}, \bibinfo
  {author} {\bibfnamefont {S.~D.}\ \bibnamefont {Jenkins}}, \ and\ \bibinfo
  {author} {\bibfnamefont {J.}~\bibnamefont {Ruostekoski}},\ }\href {\doibase
  10.1103/PhysRevLett.113.133602} {\bibfield  {journal} {\bibinfo  {journal}
  {Physical Review Letters}\ }\textbf {\bibinfo {volume} {113}},\ \bibinfo
  {pages} {133602} (\bibinfo {year} {2014})}\BibitemShut {NoStop}%
\bibitem [{\citenamefont {Jennewein}\ \emph {et~al.}(2016)\citenamefont
  {Jennewein}, \citenamefont {Besbes}, \citenamefont {Schilder}, \citenamefont
  {Jenkins}, \citenamefont {Sauvan}, \citenamefont {Ruostekoski}, \citenamefont
  {Greffet}, \citenamefont {Sortais},\ and\ \citenamefont
  {Browaeys}}]{Jennewein2016}%
  \BibitemOpen
  \bibfield  {author} {\bibinfo {author} {\bibfnamefont {S.}~\bibnamefont
  {Jennewein}}, \bibinfo {author} {\bibfnamefont {M.}~\bibnamefont {Besbes}},
  \bibinfo {author} {\bibfnamefont {N.~J.}\ \bibnamefont {Schilder}}, \bibinfo
  {author} {\bibfnamefont {S.~D.}\ \bibnamefont {Jenkins}}, \bibinfo {author}
  {\bibfnamefont {C.}~\bibnamefont {Sauvan}}, \bibinfo {author} {\bibfnamefont
  {J.}~\bibnamefont {Ruostekoski}}, \bibinfo {author} {\bibfnamefont {J.-J.}\
  \bibnamefont {Greffet}}, \bibinfo {author} {\bibfnamefont {Y.~R.~P.}\
  \bibnamefont {Sortais}}, \ and\ \bibinfo {author} {\bibfnamefont
  {A.}~\bibnamefont {Browaeys}},\ }\href {\doibase
  10.1103/PhysRevLett.116.233601} {\bibfield  {journal} {\bibinfo  {journal}
  {Physical Review Letters}\ }\textbf {\bibinfo {volume} {116}},\ \bibinfo
  {pages} {233601} (\bibinfo {year} {2016})}\BibitemShut {NoStop}%
\bibitem [{\citenamefont {Glicenstein}\ \emph {et~al.}(2020)\citenamefont
  {Glicenstein}, \citenamefont {Ferioli}, \citenamefont {{\v{S}}ibali{\'{c}}},
  \citenamefont {Brossard}, \citenamefont {Ferrier-Barbut},\ and\ \citenamefont
  {Browaeys}}]{Glicenstein2020}%
  \BibitemOpen
  \bibfield  {author} {\bibinfo {author} {\bibfnamefont {A.}~\bibnamefont
  {Glicenstein}}, \bibinfo {author} {\bibfnamefont {G.}~\bibnamefont
  {Ferioli}}, \bibinfo {author} {\bibfnamefont {N.}~\bibnamefont
  {{\v{S}}ibali{\'{c}}}}, \bibinfo {author} {\bibfnamefont {L.}~\bibnamefont
  {Brossard}}, \bibinfo {author} {\bibfnamefont {I.}~\bibnamefont
  {Ferrier-Barbut}}, \ and\ \bibinfo {author} {\bibfnamefont {A.}~\bibnamefont
  {Browaeys}},\ }\href {\doibase 10.1103/PhysRevLett.124.253602} {\bibfield
  {journal} {\bibinfo  {journal} {Physical Review Letters}\ }\textbf {\bibinfo
  {volume} {124}},\ \bibinfo {pages} {253602} (\bibinfo {year}
  {2020})}\BibitemShut {NoStop}%
\bibitem [{\citenamefont {Meir}\ \emph {et~al.}(2014)\citenamefont {Meir},
  \citenamefont {Schwartz}, \citenamefont {Shahmoon}, \citenamefont {Oron},\
  and\ \citenamefont {Ozeri}}]{Meir2014}%
  \BibitemOpen
  \bibfield  {author} {\bibinfo {author} {\bibfnamefont {Z.}~\bibnamefont
  {Meir}}, \bibinfo {author} {\bibfnamefont {O.}~\bibnamefont {Schwartz}},
  \bibinfo {author} {\bibfnamefont {E.}~\bibnamefont {Shahmoon}}, \bibinfo
  {author} {\bibfnamefont {D.}~\bibnamefont {Oron}}, \ and\ \bibinfo {author}
  {\bibfnamefont {R.}~\bibnamefont {Ozeri}},\ }\href {\doibase
  10.1103/PhysRevLett.113.193002} {\bibfield  {journal} {\bibinfo  {journal}
  {Physical Review Letters}\ }\textbf {\bibinfo {volume} {113}},\ \bibinfo
  {pages} {193002} (\bibinfo {year} {2014})}\BibitemShut {NoStop}%
\bibitem [{\citenamefont {Scheibner}\ \emph {et~al.}(2007)\citenamefont
  {Scheibner}, \citenamefont {Schmidt}, \citenamefont {Worschech},
  \citenamefont {Forchel}, \citenamefont {Bacher}, \citenamefont {Passow},\
  and\ \citenamefont {Hommel}}]{Scheibner2007}%
  \BibitemOpen
  \bibfield  {author} {\bibinfo {author} {\bibfnamefont {M.}~\bibnamefont
  {Scheibner}}, \bibinfo {author} {\bibfnamefont {T.}~\bibnamefont {Schmidt}},
  \bibinfo {author} {\bibfnamefont {L.}~\bibnamefont {Worschech}}, \bibinfo
  {author} {\bibfnamefont {A.}~\bibnamefont {Forchel}}, \bibinfo {author}
  {\bibfnamefont {G.}~\bibnamefont {Bacher}}, \bibinfo {author} {\bibfnamefont
  {T.}~\bibnamefont {Passow}}, \ and\ \bibinfo {author} {\bibfnamefont
  {D.}~\bibnamefont {Hommel}},\ }\href {\doibase 10.1038/nphys494} {\bibfield
  {journal} {\bibinfo  {journal} {Nature Physics}\ }\textbf {\bibinfo {volume}
  {3}},\ \bibinfo {pages} {106} (\bibinfo {year} {2007})}\BibitemShut {NoStop}%
\bibitem [{\citenamefont {Tighineanu}\ \emph {et~al.}(2016)\citenamefont
  {Tighineanu}, \citenamefont {Daveau}, \citenamefont {Lehmann}, \citenamefont
  {Beere}, \citenamefont {Ritchie}, \citenamefont {Lodahl},\ and\ \citenamefont
  {Stobbe}}]{Tighineanu2016}%
  \BibitemOpen
  \bibfield  {author} {\bibinfo {author} {\bibfnamefont {P.}~\bibnamefont
  {Tighineanu}}, \bibinfo {author} {\bibfnamefont {R.~S.}\ \bibnamefont
  {Daveau}}, \bibinfo {author} {\bibfnamefont {T.~B.}\ \bibnamefont {Lehmann}},
  \bibinfo {author} {\bibfnamefont {H.~E.}\ \bibnamefont {Beere}}, \bibinfo
  {author} {\bibfnamefont {D.~A.}\ \bibnamefont {Ritchie}}, \bibinfo {author}
  {\bibfnamefont {P.}~\bibnamefont {Lodahl}}, \ and\ \bibinfo {author}
  {\bibfnamefont {S.}~\bibnamefont {Stobbe}},\ }\href {\doibase
  10.1103/PhysRevLett.116.163604} {\bibfield  {journal} {\bibinfo  {journal}
  {Physical Review Letters}\ }\textbf {\bibinfo {volume} {116}},\ \bibinfo
  {pages} {163604} (\bibinfo {year} {2016})}\BibitemShut {NoStop}%
\bibitem [{\citenamefont {Mlynek}\ \emph {et~al.}(2014)\citenamefont {Mlynek},
  \citenamefont {Abdumalikov}, \citenamefont {Eichler},\ and\ \citenamefont
  {Wallraff}}]{mlynek2014observation}%
  \BibitemOpen
  \bibfield  {author} {\bibinfo {author} {\bibfnamefont {J.~A.}\ \bibnamefont
  {Mlynek}}, \bibinfo {author} {\bibfnamefont {A.~A.}\ \bibnamefont
  {Abdumalikov}}, \bibinfo {author} {\bibfnamefont {C.}~\bibnamefont
  {Eichler}}, \ and\ \bibinfo {author} {\bibfnamefont {A.}~\bibnamefont
  {Wallraff}},\ }\href@noop {} {\bibfield  {journal} {\bibinfo  {journal}
  {Nature communications}\ }\textbf {\bibinfo {volume} {5}},\ \bibinfo {pages}
  {1} (\bibinfo {year} {2014})}\BibitemShut {NoStop}%
\bibitem [{\citenamefont {Lambert}\ \emph {et~al.}(2016)\citenamefont
  {Lambert}, \citenamefont {Matsuzaki}, \citenamefont {Kakuyanagi},
  \citenamefont {Ishida}, \citenamefont {Saito},\ and\ \citenamefont
  {Nori}}]{PhysRevB.94.224510}%
  \BibitemOpen
  \bibfield  {author} {\bibinfo {author} {\bibfnamefont {N.}~\bibnamefont
  {Lambert}}, \bibinfo {author} {\bibfnamefont {Y.}~\bibnamefont {Matsuzaki}},
  \bibinfo {author} {\bibfnamefont {K.}~\bibnamefont {Kakuyanagi}}, \bibinfo
  {author} {\bibfnamefont {N.}~\bibnamefont {Ishida}}, \bibinfo {author}
  {\bibfnamefont {S.}~\bibnamefont {Saito}}, \ and\ \bibinfo {author}
  {\bibfnamefont {F.}~\bibnamefont {Nori}},\ }\href {\doibase
  10.1103/PhysRevB.94.224510} {\bibfield  {journal} {\bibinfo  {journal} {Phys.
  Rev. B}\ }\textbf {\bibinfo {volume} {94}},\ \bibinfo {pages} {224510}
  (\bibinfo {year} {2016})}\BibitemShut {NoStop}%
\bibitem [{\citenamefont {Goban}\ \emph {et~al.}(2015)\citenamefont {Goban},
  \citenamefont {Hung}, \citenamefont {Hood}, \citenamefont {Yu}, \citenamefont
  {Muniz}, \citenamefont {Painter},\ and\ \citenamefont {Kimble}}]{Goban2015}%
  \BibitemOpen
  \bibfield  {author} {\bibinfo {author} {\bibfnamefont {A.}~\bibnamefont
  {Goban}}, \bibinfo {author} {\bibfnamefont {C.-L.}\ \bibnamefont {Hung}},
  \bibinfo {author} {\bibfnamefont {J.~D.}\ \bibnamefont {Hood}}, \bibinfo
  {author} {\bibfnamefont {S.-P.}\ \bibnamefont {Yu}}, \bibinfo {author}
  {\bibfnamefont {J.~A.}\ \bibnamefont {Muniz}}, \bibinfo {author}
  {\bibfnamefont {O.}~\bibnamefont {Painter}}, \ and\ \bibinfo {author}
  {\bibfnamefont {H.~J.}\ \bibnamefont {Kimble}},\ }\href {\doibase
  10.1103/PhysRevLett.115.063601} {\bibfield  {journal} {\bibinfo  {journal}
  {Physical Review Letters}\ }\textbf {\bibinfo {volume} {115}},\ \bibinfo
  {pages} {063601} (\bibinfo {year} {2015})}\BibitemShut {NoStop}%
\bibitem [{\citenamefont {Asenjo-Garcia}\ \emph {et~al.}(2017)\citenamefont
  {Asenjo-Garcia}, \citenamefont {Hood}, \citenamefont {Chang},\ and\
  \citenamefont {Kimble}}]{Asenjo-Garcia2017}%
  \BibitemOpen
  \bibfield  {author} {\bibinfo {author} {\bibfnamefont {A.}~\bibnamefont
  {Asenjo-Garcia}}, \bibinfo {author} {\bibfnamefont {J.~D.}\ \bibnamefont
  {Hood}}, \bibinfo {author} {\bibfnamefont {D.~E.}\ \bibnamefont {Chang}}, \
  and\ \bibinfo {author} {\bibfnamefont {H.~J.}\ \bibnamefont {Kimble}},\
  }\href {\doibase 10.1103/PhysRevA.95.033818} {\bibfield  {journal} {\bibinfo
  {journal} {Physical Review A}\ }\textbf {\bibinfo {volume} {95}},\ \bibinfo
  {pages} {033818} (\bibinfo {year} {2017})}\BibitemShut {NoStop}%
\bibitem [{\citenamefont {Petrosyan}\ and\ \citenamefont
  {Kurizki}(2002)}]{Petrosyan2002}%
  \BibitemOpen
  \bibfield  {author} {\bibinfo {author} {\bibfnamefont {D.}~\bibnamefont
  {Petrosyan}}\ and\ \bibinfo {author} {\bibfnamefont {G.}~\bibnamefont
  {Kurizki}},\ }\href {\doibase 10.1103/PhysRevLett.89.207902} {\bibfield
  {journal} {\bibinfo  {journal} {Physical Review Letters}\ }\textbf {\bibinfo
  {volume} {89}},\ \bibinfo {pages} {207902} (\bibinfo {year}
  {2002})}\BibitemShut {NoStop}%
\bibitem [{\citenamefont {Scully}\ \emph {et~al.}(2006)\citenamefont {Scully},
  \citenamefont {Fry}, \citenamefont {Ooi},\ and\ \citenamefont
  {W{\'{o}}dkiewicz}}]{Scully2006}%
  \BibitemOpen
  \bibfield  {author} {\bibinfo {author} {\bibfnamefont {M.~O.}\ \bibnamefont
  {Scully}}, \bibinfo {author} {\bibfnamefont {E.~S.}\ \bibnamefont {Fry}},
  \bibinfo {author} {\bibfnamefont {C.~H.~R.}\ \bibnamefont {Ooi}}, \ and\
  \bibinfo {author} {\bibfnamefont {K.}~\bibnamefont {W{\'{o}}dkiewicz}},\
  }\href {\doibase 10.1103/PhysRevLett.96.010501} {\bibfield  {journal}
  {\bibinfo  {journal} {Physical Review Letters}\ }\textbf {\bibinfo {volume}
  {96}},\ \bibinfo {pages} {010501} (\bibinfo {year} {2006})}\BibitemShut
  {NoStop}%
\bibitem [{\citenamefont {Scully}\ and\ \citenamefont
  {Svidzinsky}(2009)}]{Scully2009}%
  \BibitemOpen
  \bibfield  {author} {\bibinfo {author} {\bibfnamefont {M.~O.}\ \bibnamefont
  {Scully}}\ and\ \bibinfo {author} {\bibfnamefont {A.~A.}\ \bibnamefont
  {Svidzinsky}},\ }\href {\doibase 10.1126/science.1176695} {\bibfield
  {journal} {\bibinfo  {journal} {Science}\ }\textbf {\bibinfo {volume}
  {325}},\ \bibinfo {pages} {1510} (\bibinfo {year} {2009})}\BibitemShut
  {NoStop}%
\bibitem [{\citenamefont {Scully}(2009)}]{Scully2009a}%
  \BibitemOpen
  \bibfield  {author} {\bibinfo {author} {\bibfnamefont {M.~O.}\ \bibnamefont
  {Scully}},\ }\href {\doibase 10.1103/PhysRevLett.102.143601} {\bibfield
  {journal} {\bibinfo  {journal} {Physical Review Letters}\ }\textbf {\bibinfo
  {volume} {102}},\ \bibinfo {pages} {143601} (\bibinfo {year}
  {2009})}\BibitemShut {NoStop}%
\bibitem [{\citenamefont {Manassah}(2009)}]{Manassah2009}%
  \BibitemOpen
  \bibfield  {author} {\bibinfo {author} {\bibfnamefont {J.~T.}\ \bibnamefont
  {Manassah}},\ }\href {\doibase 10.1134/S1054660X09210087} {\bibfield
  {journal} {\bibinfo  {journal} {Laser Physics}\ }\textbf {\bibinfo {volume}
  {19}},\ \bibinfo {pages} {2102} (\bibinfo {year} {2009})}\BibitemShut
  {NoStop}%
\bibitem [{\citenamefont {Mazets}\ and\ \citenamefont
  {Kurizki}(2007)}]{Mazets2007}%
  \BibitemOpen
  \bibfield  {author} {\bibinfo {author} {\bibfnamefont {I.~E.}\ \bibnamefont
  {Mazets}}\ and\ \bibinfo {author} {\bibfnamefont {G.}~\bibnamefont
  {Kurizki}},\ }\href {\doibase 10.1088/0953-4075/40/6/F01} {\bibfield
  {journal} {\bibinfo  {journal} {Journal of Physics B: Atomic, Molecular and
  Optical Physics}\ }\textbf {\bibinfo {volume} {40}},\ \bibinfo {pages} {F105}
  (\bibinfo {year} {2007})}\BibitemShut {NoStop}%
\bibitem [{\citenamefont {Friedberg}\ and\ \citenamefont
  {Manassah}(2008)}]{Friedberg2008}%
  \BibitemOpen
  \bibfield  {author} {\bibinfo {author} {\bibfnamefont {R.}~\bibnamefont
  {Friedberg}}\ and\ \bibinfo {author} {\bibfnamefont {J.~T.}\ \bibnamefont
  {Manassah}},\ }\href {\doibase 10.1016/j.physleta.2007.11.064} {\bibfield
  {journal} {\bibinfo  {journal} {Physics Letters A}\ }\textbf {\bibinfo
  {volume} {372}},\ \bibinfo {pages} {2514} (\bibinfo {year}
  {2008})}\BibitemShut {NoStop}%
\bibitem [{\citenamefont {Mirza}\ and\ \citenamefont
  {Begzjav}(2016)}]{Mirza2016}%
  \BibitemOpen
  \bibfield  {author} {\bibinfo {author} {\bibfnamefont {I.~M.}\ \bibnamefont
  {Mirza}}\ and\ \bibinfo {author} {\bibfnamefont {T.}~\bibnamefont
  {Begzjav}},\ }\href {\doibase 10.1209/0295-5075/114/24004} {\bibfield
  {journal} {\bibinfo  {journal} {EPL (Europhysics Letters)}\ }\textbf
  {\bibinfo {volume} {114}},\ \bibinfo {pages} {24004} (\bibinfo {year}
  {2016})}\BibitemShut {NoStop}%
\bibitem [{\citenamefont {Paris-Mandoki}\ \emph {et~al.}(2017)\citenamefont
  {Paris-Mandoki}, \citenamefont {Braun}, \citenamefont {Kumlin}, \citenamefont
  {Tresp}, \citenamefont {Mirgorodskiy}, \citenamefont {Christaller},
  \citenamefont {B{\"{u}}chler},\ and\ \citenamefont
  {Hofferberth}}]{Paris-Mandoki2017}%
  \BibitemOpen
  \bibfield  {author} {\bibinfo {author} {\bibfnamefont {A.}~\bibnamefont
  {Paris-Mandoki}}, \bibinfo {author} {\bibfnamefont {C.}~\bibnamefont
  {Braun}}, \bibinfo {author} {\bibfnamefont {J.}~\bibnamefont {Kumlin}},
  \bibinfo {author} {\bibfnamefont {C.}~\bibnamefont {Tresp}}, \bibinfo
  {author} {\bibfnamefont {I.}~\bibnamefont {Mirgorodskiy}}, \bibinfo {author}
  {\bibfnamefont {F.}~\bibnamefont {Christaller}}, \bibinfo {author}
  {\bibfnamefont {H.~P.}\ \bibnamefont {B{\"{u}}chler}}, \ and\ \bibinfo
  {author} {\bibfnamefont {S.}~\bibnamefont {Hofferberth}},\ }\href {\doibase
  10.1103/PhysRevX.7.041010} {\bibfield  {journal} {\bibinfo  {journal}
  {Physical Review X}\ }\textbf {\bibinfo {volume} {7}},\ \bibinfo {pages}
  {041010} (\bibinfo {year} {2017})}\BibitemShut {NoStop}%
\bibitem [{\citenamefont {Guerin}\ \emph {et~al.}(2016)\citenamefont {Guerin},
  \citenamefont {Ara{\'{u}}jo},\ and\ \citenamefont {Kaiser}}]{Guerin2016}%
  \BibitemOpen
  \bibfield  {author} {\bibinfo {author} {\bibfnamefont {W.}~\bibnamefont
  {Guerin}}, \bibinfo {author} {\bibfnamefont {M.~O.}\ \bibnamefont
  {Ara{\'{u}}jo}}, \ and\ \bibinfo {author} {\bibfnamefont {R.}~\bibnamefont
  {Kaiser}},\ }\href {\doibase 10.1103/PhysRevLett.116.083601} {\bibfield
  {journal} {\bibinfo  {journal} {Physical Review Letters}\ }\textbf {\bibinfo
  {volume} {116}},\ \bibinfo {pages} {083601} (\bibinfo {year}
  {2016})}\BibitemShut {NoStop}%
\bibitem [{\citenamefont {Bettles}\ \emph {et~al.}(2018)\citenamefont
  {Bettles}, \citenamefont {Ilieva}, \citenamefont {Busche}, \citenamefont
  {Huillery}, \citenamefont {Ball}, \citenamefont {Spong},\ and\ \citenamefont
  {Adams}}]{Bettles2018}%
  \BibitemOpen
  \bibfield  {author} {\bibinfo {author} {\bibfnamefont {R.~J.}\ \bibnamefont
  {Bettles}}, \bibinfo {author} {\bibfnamefont {T.}~\bibnamefont {Ilieva}},
  \bibinfo {author} {\bibfnamefont {H.}~\bibnamefont {Busche}}, \bibinfo
  {author} {\bibfnamefont {P.}~\bibnamefont {Huillery}}, \bibinfo {author}
  {\bibfnamefont {S.~W.}\ \bibnamefont {Ball}}, \bibinfo {author}
  {\bibfnamefont {N.~L.~R.}\ \bibnamefont {Spong}}, \ and\ \bibinfo {author}
  {\bibfnamefont {C.~S.}\ \bibnamefont {Adams}},\ }\href
  {http://arxiv.org/abs/1808.08415} {\  (\bibinfo {year} {2018})},\ \Eprint
  {http://arxiv.org/abs/1808.08415} {arXiv:1808.08415} \BibitemShut {NoStop}%
\bibitem [{\citenamefont {Lehmberg}(1970)}]{Lehmberg1970}%
  \BibitemOpen
  \bibfield  {author} {\bibinfo {author} {\bibfnamefont {R.~H.}\ \bibnamefont
  {Lehmberg}},\ }\href {\doibase 10.1103/PhysRevA.2.883} {\bibfield  {journal}
  {\bibinfo  {journal} {Physical Review A}\ }\textbf {\bibinfo {volume} {2}},\
  \bibinfo {pages} {883} (\bibinfo {year} {1970})}\BibitemShut {NoStop}%
\bibitem [{\citenamefont {Kumlin}\ \emph {et~al.}(2018)\citenamefont {Kumlin},
  \citenamefont {Hofferberth},\ and\ \citenamefont
  {B{\"{u}}chler}}]{Kumlin2018}%
  \BibitemOpen
  \bibfield  {author} {\bibinfo {author} {\bibfnamefont {J.}~\bibnamefont
  {Kumlin}}, \bibinfo {author} {\bibfnamefont {S.}~\bibnamefont {Hofferberth}},
  \ and\ \bibinfo {author} {\bibfnamefont {H.~P.}\ \bibnamefont
  {B{\"{u}}chler}},\ }\href {\doibase 10.1103/PhysRevLett.121.013601}
  {\bibfield  {journal} {\bibinfo  {journal} {Physical Review Letters}\
  }\textbf {\bibinfo {volume} {121}},\ \bibinfo {pages} {013601} (\bibinfo
  {year} {2018})}\BibitemShut {NoStop}%
\bibitem [{\citenamefont {Svidzinsky}\ \emph {et~al.}(2010)\citenamefont
  {Svidzinsky}, \citenamefont {Chang},\ and\ \citenamefont
  {Scully}}]{Svidzinsky2010}%
  \BibitemOpen
  \bibfield  {author} {\bibinfo {author} {\bibfnamefont {A.~A.}\ \bibnamefont
  {Svidzinsky}}, \bibinfo {author} {\bibfnamefont {J.-T.}\ \bibnamefont
  {Chang}}, \ and\ \bibinfo {author} {\bibfnamefont {M.~O.}\ \bibnamefont
  {Scully}},\ }\href {\doibase 10.1103/PhysRevA.81.053821} {\bibfield
  {journal} {\bibinfo  {journal} {Physical Review A}\ }\textbf {\bibinfo
  {volume} {81}},\ \bibinfo {pages} {053821} (\bibinfo {year}
  {2010})}\BibitemShut {NoStop}%
\bibitem [{\citenamefont {Grankin}\ \emph {et~al.}(2018)\citenamefont
  {Grankin}, \citenamefont {Guimond}, \citenamefont {Vasilyev}, \citenamefont
  {Vermersch},\ and\ \citenamefont {Zoller}}]{Grankin2018}%
  \BibitemOpen
  \bibfield  {author} {\bibinfo {author} {\bibfnamefont {A.}~\bibnamefont
  {Grankin}}, \bibinfo {author} {\bibfnamefont {P.~O.}\ \bibnamefont
  {Guimond}}, \bibinfo {author} {\bibfnamefont {D.~V.}\ \bibnamefont
  {Vasilyev}}, \bibinfo {author} {\bibfnamefont {B.}~\bibnamefont {Vermersch}},
  \ and\ \bibinfo {author} {\bibfnamefont {P.}~\bibnamefont {Zoller}},\ }\href
  {\doibase 10.1103/PhysRevA.98.043825} {\bibfield  {journal} {\bibinfo
  {journal} {Physical Review A}\ }\textbf {\bibinfo {volume} {98}},\ \bibinfo
  {pages} {043825} (\bibinfo {year} {2018})}\BibitemShut {NoStop}%
\bibitem [{\citenamefont {Chang}\ \emph {et~al.}(2012)\citenamefont {Chang},
  \citenamefont {Jiang}, \citenamefont {Gorshkov},\ and\ \citenamefont
  {Kimble}}]{Chang2012}%
  \BibitemOpen
  \bibfield  {author} {\bibinfo {author} {\bibfnamefont {D.~E.}\ \bibnamefont
  {Chang}}, \bibinfo {author} {\bibfnamefont {L.}~\bibnamefont {Jiang}},
  \bibinfo {author} {\bibfnamefont {A.~V.}\ \bibnamefont {Gorshkov}}, \ and\
  \bibinfo {author} {\bibfnamefont {H.~J.}\ \bibnamefont {Kimble}},\ }\href
  {\doibase 10.1088/1367-2630/14/6/063003} {\bibfield  {journal} {\bibinfo
  {journal} {New Journal of Physics}\ }\textbf {\bibinfo {volume} {14}},\
  \bibinfo {pages} {063003} (\bibinfo {year} {2012})}\BibitemShut {NoStop}%
\bibitem [{\citenamefont {Bettles}\ \emph {et~al.}(2016)\citenamefont
  {Bettles}, \citenamefont {Gardiner},\ and\ \citenamefont
  {Adams}}]{Bettles2016}%
  \BibitemOpen
  \bibfield  {author} {\bibinfo {author} {\bibfnamefont {R.~J.}\ \bibnamefont
  {Bettles}}, \bibinfo {author} {\bibfnamefont {S.~A.}\ \bibnamefont
  {Gardiner}}, \ and\ \bibinfo {author} {\bibfnamefont {C.~S.}\ \bibnamefont
  {Adams}},\ }\href {\doibase 10.1103/PhysRevLett.116.103602} {\bibfield
  {journal} {\bibinfo  {journal} {Physical Review Letters}\ }\textbf {\bibinfo
  {volume} {116}},\ \bibinfo {pages} {103602} (\bibinfo {year}
  {2016})}\BibitemShut {NoStop}%
\bibitem [{\citenamefont {Shahmoon}\ \emph {et~al.}(2017)\citenamefont
  {Shahmoon}, \citenamefont {Wild}, \citenamefont {Lukin},\ and\ \citenamefont
  {Yelin}}]{Shahmoon2017}%
  \BibitemOpen
  \bibfield  {author} {\bibinfo {author} {\bibfnamefont {E.}~\bibnamefont
  {Shahmoon}}, \bibinfo {author} {\bibfnamefont {D.~S.}\ \bibnamefont {Wild}},
  \bibinfo {author} {\bibfnamefont {M.~D.}\ \bibnamefont {Lukin}}, \ and\
  \bibinfo {author} {\bibfnamefont {S.~F.}\ \bibnamefont {Yelin}},\ }\href
  {\doibase 10.1103/PhysRevLett.118.113601} {\bibfield  {journal} {\bibinfo
  {journal} {Physical Review Letters}\ }\textbf {\bibinfo {volume} {118}},\
  \bibinfo {pages} {113601} (\bibinfo {year} {2017})}\BibitemShut {NoStop}%
\bibitem [{\citenamefont {Rui}\ \emph {et~al.}(2020)\citenamefont {Rui},
  \citenamefont {Wei}, \citenamefont {Rubio-Abadal}, \citenamefont {Hollerith},
  \citenamefont {Zeiher}, \citenamefont {Stamper-Kurn}, \citenamefont {Gross},\
  and\ \citenamefont {Bloch}}]{Rui2020}%
  \BibitemOpen
  \bibfield  {author} {\bibinfo {author} {\bibfnamefont {J.}~\bibnamefont
  {Rui}}, \bibinfo {author} {\bibfnamefont {D.}~\bibnamefont {Wei}}, \bibinfo
  {author} {\bibfnamefont {A.}~\bibnamefont {Rubio-Abadal}}, \bibinfo {author}
  {\bibfnamefont {S.}~\bibnamefont {Hollerith}}, \bibinfo {author}
  {\bibfnamefont {J.}~\bibnamefont {Zeiher}}, \bibinfo {author} {\bibfnamefont
  {D.~M.}\ \bibnamefont {Stamper-Kurn}}, \bibinfo {author} {\bibfnamefont
  {C.}~\bibnamefont {Gross}}, \ and\ \bibinfo {author} {\bibfnamefont
  {I.}~\bibnamefont {Bloch}},\ }\href {http://arxiv.org/abs/2001.00795} {\
  (\bibinfo {year} {2020})},\ \Eprint {http://arxiv.org/abs/2001.00795}
  {arXiv:2001.00795} \BibitemShut {NoStop}%
\bibitem [{\citenamefont {Vetsch}\ \emph {et~al.}(2010)\citenamefont {Vetsch},
  \citenamefont {Reitz}, \citenamefont {Sagu{\'{e}}}, \citenamefont {Schmidt},
  \citenamefont {Dawkins},\ and\ \citenamefont {Rauschenbeutel}}]{Vetsch2010}%
  \BibitemOpen
  \bibfield  {author} {\bibinfo {author} {\bibfnamefont {E.}~\bibnamefont
  {Vetsch}}, \bibinfo {author} {\bibfnamefont {D.}~\bibnamefont {Reitz}},
  \bibinfo {author} {\bibfnamefont {G.}~\bibnamefont {Sagu{\'{e}}}}, \bibinfo
  {author} {\bibfnamefont {R.}~\bibnamefont {Schmidt}}, \bibinfo {author}
  {\bibfnamefont {S.~T.}\ \bibnamefont {Dawkins}}, \ and\ \bibinfo {author}
  {\bibfnamefont {A.}~\bibnamefont {Rauschenbeutel}},\ }\href {\doibase
  10.1103/PhysRevLett.104.203603} {\bibfield  {journal} {\bibinfo  {journal}
  {Physical Review Letters}\ }\textbf {\bibinfo {volume} {104}},\ \bibinfo
  {pages} {203603} (\bibinfo {year} {2010})}\BibitemShut {NoStop}%
\bibitem [{\citenamefont {Solano}\ \emph {et~al.}(2017)\citenamefont {Solano},
  \citenamefont {Barberis-Blostein}, \citenamefont {Fatemi}, \citenamefont
  {Orozco},\ and\ \citenamefont {Rolston}}]{Solano2017}%
  \BibitemOpen
  \bibfield  {author} {\bibinfo {author} {\bibfnamefont {P.}~\bibnamefont
  {Solano}}, \bibinfo {author} {\bibfnamefont {P.}~\bibnamefont
  {Barberis-Blostein}}, \bibinfo {author} {\bibfnamefont {F.~K.}\ \bibnamefont
  {Fatemi}}, \bibinfo {author} {\bibfnamefont {L.~A.}\ \bibnamefont {Orozco}},
  \ and\ \bibinfo {author} {\bibfnamefont {S.~L.}\ \bibnamefont {Rolston}},\
  }\href {\doibase 10.1038/s41467-017-01994-3} {\bibfield  {journal} {\bibinfo
  {journal} {Nature Communications}\ }\textbf {\bibinfo {volume} {8}},\
  \bibinfo {pages} {1857} (\bibinfo {year} {2017})}\BibitemShut {NoStop}%
\bibitem [{\citenamefont {Lodahl}\ \emph {et~al.}(2017)\citenamefont {Lodahl},
  \citenamefont {Mahmoodian}, \citenamefont {Stobbe}, \citenamefont
  {Rauschenbeutel}, \citenamefont {Schneeweiss}, \citenamefont {Volz},
  \citenamefont {Pichler},\ and\ \citenamefont {Zoller}}]{Lodahl2017}%
  \BibitemOpen
  \bibfield  {author} {\bibinfo {author} {\bibfnamefont {P.}~\bibnamefont
  {Lodahl}}, \bibinfo {author} {\bibfnamefont {S.}~\bibnamefont {Mahmoodian}},
  \bibinfo {author} {\bibfnamefont {S.}~\bibnamefont {Stobbe}}, \bibinfo
  {author} {\bibfnamefont {A.}~\bibnamefont {Rauschenbeutel}}, \bibinfo
  {author} {\bibfnamefont {P.}~\bibnamefont {Schneeweiss}}, \bibinfo {author}
  {\bibfnamefont {J.}~\bibnamefont {Volz}}, \bibinfo {author} {\bibfnamefont
  {H.}~\bibnamefont {Pichler}}, \ and\ \bibinfo {author} {\bibfnamefont
  {P.}~\bibnamefont {Zoller}},\ }\href {\doibase 10.1038/nature21037}
  {\bibfield  {journal} {\bibinfo  {journal} {Nature}\ }\textbf {\bibinfo
  {volume} {541}},\ \bibinfo {pages} {473} (\bibinfo {year}
  {2017})}\BibitemShut {NoStop}%
\bibitem [{\citenamefont {Mahmoodian}\ \emph {et~al.}(2018)\citenamefont
  {Mahmoodian}, \citenamefont {{\v{C}}epulkovskis}, \citenamefont {Das},
  \citenamefont {Lodahl}, \citenamefont {Hammerer},\ and\ \citenamefont
  {S{\o}rensen}}]{Mahmoodian2018}%
  \BibitemOpen
  \bibfield  {author} {\bibinfo {author} {\bibfnamefont {S.}~\bibnamefont
  {Mahmoodian}}, \bibinfo {author} {\bibfnamefont {M.}~\bibnamefont
  {{\v{C}}epulkovskis}}, \bibinfo {author} {\bibfnamefont {S.}~\bibnamefont
  {Das}}, \bibinfo {author} {\bibfnamefont {P.}~\bibnamefont {Lodahl}},
  \bibinfo {author} {\bibfnamefont {K.}~\bibnamefont {Hammerer}}, \ and\
  \bibinfo {author} {\bibfnamefont {A.~S.}\ \bibnamefont {S{\o}rensen}},\
  }\href {\doibase 10.1103/PhysRevLett.121.143601} {\bibfield  {journal}
  {\bibinfo  {journal} {Physical Review Letters}\ }\textbf {\bibinfo {volume}
  {121}},\ \bibinfo {pages} {143601} (\bibinfo {year} {2018})}\BibitemShut
  {NoStop}%
\bibitem [{\citenamefont {Olmos}\ \emph {et~al.}(2020)\citenamefont {Olmos},
  \citenamefont {Buonaiuto}, \citenamefont {Schneeweiss},\ and\ \citenamefont
  {Lesanovsky}}]{Olmos2020}%
  \BibitemOpen
  \bibfield  {author} {\bibinfo {author} {\bibfnamefont {B.}~\bibnamefont
  {Olmos}}, \bibinfo {author} {\bibfnamefont {G.}~\bibnamefont {Buonaiuto}},
  \bibinfo {author} {\bibfnamefont {P.}~\bibnamefont {Schneeweiss}}, \ and\
  \bibinfo {author} {\bibfnamefont {I.}~\bibnamefont {Lesanovsky}},\ }\href
  {http://arxiv.org/abs/2003.01620} {\  (\bibinfo {year} {2020})},\ \Eprint
  {http://arxiv.org/abs/2003.01620} {arXiv:2003.01620} \BibitemShut {NoStop}%
\bibitem [{\citenamefont {Iversen}\ and\ \citenamefont
  {Pohl}(2020)}]{Iversen2020}%
  \BibitemOpen
  \bibfield  {author} {\bibinfo {author} {\bibfnamefont {O.~A.}\ \bibnamefont
  {Iversen}}\ and\ \bibinfo {author} {\bibfnamefont {T.}~\bibnamefont {Pohl}},\
  }\href {http://arxiv.org/abs/2006.03408} {\  (\bibinfo {year} {2020})},\
  \Eprint {http://arxiv.org/abs/2006.03408} {arXiv:2006.03408} \BibitemShut
  {NoStop}%
\bibitem [{\citenamefont {Shi}\ \emph {et~al.}(2015)\citenamefont {Shi},
  \citenamefont {Chang},\ and\ \citenamefont {Cirac}}]{Shi2015}%
  \BibitemOpen
  \bibfield  {author} {\bibinfo {author} {\bibfnamefont {T.}~\bibnamefont
  {Shi}}, \bibinfo {author} {\bibfnamefont {D.~E.}\ \bibnamefont {Chang}}, \
  and\ \bibinfo {author} {\bibfnamefont {J.~I.}\ \bibnamefont {Cirac}},\ }\href
  {\doibase 10.1103/PhysRevA.92.053834} {\bibfield  {journal} {\bibinfo
  {journal} {Physical Review A}\ }\textbf {\bibinfo {volume} {92}},\ \bibinfo
  {pages} {053834} (\bibinfo {year} {2015})}\BibitemShut {NoStop}%
\bibitem [{\citenamefont {Pichler}\ \emph {et~al.}(2015)\citenamefont
  {Pichler}, \citenamefont {Ramos}, \citenamefont {Daley},\ and\ \citenamefont
  {Zoller}}]{Pichler2015}%
  \BibitemOpen
  \bibfield  {author} {\bibinfo {author} {\bibfnamefont {H.}~\bibnamefont
  {Pichler}}, \bibinfo {author} {\bibfnamefont {T.}~\bibnamefont {Ramos}},
  \bibinfo {author} {\bibfnamefont {A.~J.}\ \bibnamefont {Daley}}, \ and\
  \bibinfo {author} {\bibfnamefont {P.}~\bibnamefont {Zoller}},\ }\href
  {\doibase 10.1103/PhysRevA.91.042116} {\bibfield  {journal} {\bibinfo
  {journal} {Physical Review A}\ }\textbf {\bibinfo {volume} {91}},\ \bibinfo
  {pages} {042116} (\bibinfo {year} {2015})}\BibitemShut {NoStop}%
\bibitem [{\citenamefont {Ruostekoski}\ and\ \citenamefont
  {Javanainen}(2016)}]{Ruostekoski2016a}%
  \BibitemOpen
  \bibfield  {author} {\bibinfo {author} {\bibfnamefont {J.}~\bibnamefont
  {Ruostekoski}}\ and\ \bibinfo {author} {\bibfnamefont {J.}~\bibnamefont
  {Javanainen}},\ }\href {\doibase 10.1103/PhysRevLett.117.143602} {\bibfield
  {journal} {\bibinfo  {journal} {Physical Review Letters}\ }\textbf {\bibinfo
  {volume} {117}},\ \bibinfo {pages} {143602} (\bibinfo {year}
  {2016})}\BibitemShut {NoStop}%
\bibitem [{\citenamefont {{Le Kien}}\ and\ \citenamefont
  {Rauschenbeutel}(2017)}]{LeKien2017}%
  \BibitemOpen
  \bibfield  {author} {\bibinfo {author} {\bibfnamefont {F.}~\bibnamefont {{Le
  Kien}}}\ and\ \bibinfo {author} {\bibfnamefont {A.}~\bibnamefont
  {Rauschenbeutel}},\ }\href {\doibase 10.1103/PhysRevA.95.023838} {\bibfield
  {journal} {\bibinfo  {journal} {Physical Review A}\ }\textbf {\bibinfo
  {volume} {95}},\ \bibinfo {pages} {023838} (\bibinfo {year}
  {2017})}\BibitemShut {NoStop}%
\bibitem [{\citenamefont {Berman}(2020)}]{Berman2020}%
  \BibitemOpen
  \bibfield  {author} {\bibinfo {author} {\bibfnamefont {P.~R.}\ \bibnamefont
  {Berman}},\ }\href {\doibase 10.1103/PhysRevA.101.013830} {\bibfield
  {journal} {\bibinfo  {journal} {Physical Review A}\ }\textbf {\bibinfo
  {volume} {101}},\ \bibinfo {pages} {013830} (\bibinfo {year}
  {2020})}\BibitemShut {NoStop}%
\bibitem [{\citenamefont {Jen}\ \emph {et~al.}(2020)\citenamefont {Jen},
  \citenamefont {Chang}, \citenamefont {Lin},\ and\ \citenamefont
  {Chen}}]{Jen2020}%
  \BibitemOpen
  \bibfield  {author} {\bibinfo {author} {\bibfnamefont {H.~H.}\ \bibnamefont
  {Jen}}, \bibinfo {author} {\bibfnamefont {M.-S.}\ \bibnamefont {Chang}},
  \bibinfo {author} {\bibfnamefont {G.-D.}\ \bibnamefont {Lin}}, \ and\
  \bibinfo {author} {\bibfnamefont {Y.-C.}\ \bibnamefont {Chen}},\ }\href
  {\doibase 10.1103/PhysRevA.101.023830} {\bibfield  {journal} {\bibinfo
  {journal} {Physical Review A}\ }\textbf {\bibinfo {volume} {101}},\ \bibinfo
  {pages} {023830} (\bibinfo {year} {2020})},\ \Eprint
  {http://arxiv.org/abs/1905.00558} {arXiv:1905.00558} \BibitemShut {NoStop}%
\bibitem [{com()}]{comment}%
  \BibitemOpen
  \href@noop {} {\bibinfo  {journal} {Note that in the limit $k \sigma \to
  \infty$, the precise form of the distribution does not matter for the final
  result. Different distribution only give rise to different corrections for
  finite $k \sigma$. A Gaussian distribution, for example, leads to exponential
  corrections in $k \sigma$, while a uniform distribution (with width $\sigma$)
  gives rise to corrections $\sim 1 / (k \sigma)^2$.}\ }\BibitemShut {NoStop}%
\bibitem [{\citenamefont {Gardiner}(1993)}]{Gardiner1993}%
  \BibitemOpen
\bibfield  {journal} {  }\bibfield  {author} {\bibinfo {author} {\bibfnamefont
  {C.~W.}\ \bibnamefont {Gardiner}},\ }\href {\doibase
  10.1103/PhysRevLett.70.2269} {\bibfield  {journal} {\bibinfo  {journal}
  {Physical Review Letters}\ }\textbf {\bibinfo {volume} {70}},\ \bibinfo
  {pages} {2269} (\bibinfo {year} {1993})}\BibitemShut {NoStop}%
\bibitem [{\citenamefont {Carmichael}(1993)}]{Carmichael1993}%
  \BibitemOpen
  \bibfield  {author} {\bibinfo {author} {\bibfnamefont {H.~J.}\ \bibnamefont
  {Carmichael}},\ }\href {\doibase 10.1103/PhysRevLett.70.2273} {\bibfield
  {journal} {\bibinfo  {journal} {Physical Review Letters}\ }\textbf {\bibinfo
  {volume} {70}},\ \bibinfo {pages} {2273} (\bibinfo {year}
  {1993})}\BibitemShut {NoStop}%
\bibitem [{\citenamefont {Stannigel}\ \emph {et~al.}(2010)\citenamefont
  {Stannigel}, \citenamefont {Rabl}, \citenamefont {S{\o}rensen}, \citenamefont
  {Zoller},\ and\ \citenamefont {Lukin}}]{Stannigel2010}%
  \BibitemOpen
  \bibfield  {author} {\bibinfo {author} {\bibfnamefont {K.}~\bibnamefont
  {Stannigel}}, \bibinfo {author} {\bibfnamefont {P.}~\bibnamefont {Rabl}},
  \bibinfo {author} {\bibfnamefont {A.~S.}\ \bibnamefont {S{\o}rensen}},
  \bibinfo {author} {\bibfnamefont {P.}~\bibnamefont {Zoller}}, \ and\ \bibinfo
  {author} {\bibfnamefont {M.~D.}\ \bibnamefont {Lukin}},\ }\href {\doibase
  10.1103/PhysRevLett.105.220501} {\bibfield  {journal} {\bibinfo  {journal}
  {Physical Review Letters}\ }\textbf {\bibinfo {volume} {105}},\ \bibinfo
  {pages} {220501} (\bibinfo {year} {2010})}\BibitemShut {NoStop}%
\bibitem [{\citenamefont {R{\"{o}}hlsberger}(2013)}]{Rohlsberger2013}%
  \BibitemOpen
  \bibfield  {author} {\bibinfo {author} {\bibfnamefont {R.}~\bibnamefont
  {R{\"{o}}hlsberger}},\ }\href {\doibase 10.1002/prop.201200074} {\bibfield
  {journal} {\bibinfo  {journal} {Fortschritte der Physik}\ }\textbf {\bibinfo
  {volume} {61}},\ \bibinfo {pages} {360} (\bibinfo {year} {2013})}\BibitemShut
  {NoStop}%
\bibitem [{\citenamefont {Petersen}\ \emph {et~al.}(2014)\citenamefont
  {Petersen}, \citenamefont {Volz},\ and\ \citenamefont
  {Rauschenbeutel}}]{Petersen2014}%
  \BibitemOpen
  \bibfield  {author} {\bibinfo {author} {\bibfnamefont {J.}~\bibnamefont
  {Petersen}}, \bibinfo {author} {\bibfnamefont {J.}~\bibnamefont {Volz}}, \
  and\ \bibinfo {author} {\bibfnamefont {A.}~\bibnamefont {Rauschenbeutel}},\
  }\href {\doibase 10.1126/science.1257671} {\bibfield  {journal} {\bibinfo
  {journal} {Science}\ }\textbf {\bibinfo {volume} {346}},\ \bibinfo {pages}
  {67} (\bibinfo {year} {2014})}\BibitemShut {NoStop}%
\bibitem [{\citenamefont {Hood}\ \emph {et~al.}(2016)\citenamefont {Hood},
  \citenamefont {Goban}, \citenamefont {Asenjo-Garcia}, \citenamefont {Lu},
  \citenamefont {Yu}, \citenamefont {Chang},\ and\ \citenamefont
  {Kimble}}]{Hood2016}%
  \BibitemOpen
  \bibfield  {author} {\bibinfo {author} {\bibfnamefont {J.~D.}\ \bibnamefont
  {Hood}}, \bibinfo {author} {\bibfnamefont {A.}~\bibnamefont {Goban}},
  \bibinfo {author} {\bibfnamefont {A.}~\bibnamefont {Asenjo-Garcia}}, \bibinfo
  {author} {\bibfnamefont {M.}~\bibnamefont {Lu}}, \bibinfo {author}
  {\bibfnamefont {S.-P.}\ \bibnamefont {Yu}}, \bibinfo {author} {\bibfnamefont
  {D.~E.}\ \bibnamefont {Chang}}, \ and\ \bibinfo {author} {\bibfnamefont
  {H.~J.}\ \bibnamefont {Kimble}},\ }\href {\doibase 10.1073/pnas.1603788113}
  {\bibfield  {journal} {\bibinfo  {journal} {Proceedings of the National
  Academy of Sciences}\ }\textbf {\bibinfo {volume} {113}},\ \bibinfo {pages}
  {10507} (\bibinfo {year} {2016})}\BibitemShut {NoStop}%
\bibitem [{\citenamefont {Lodahl}\ \emph {et~al.}(2004)\citenamefont {Lodahl},
  \citenamefont {{Floris van Driel}}, \citenamefont {Nikolaev}, \citenamefont
  {Irman}, \citenamefont {Overgaag}, \citenamefont {Vanmaekelbergh},\ and\
  \citenamefont {Vos}}]{Lodahl2004}%
  \BibitemOpen
  \bibfield  {author} {\bibinfo {author} {\bibfnamefont {P.}~\bibnamefont
  {Lodahl}}, \bibinfo {author} {\bibfnamefont {A.}~\bibnamefont {{Floris van
  Driel}}}, \bibinfo {author} {\bibfnamefont {I.~S.}\ \bibnamefont {Nikolaev}},
  \bibinfo {author} {\bibfnamefont {A.}~\bibnamefont {Irman}}, \bibinfo
  {author} {\bibfnamefont {K.}~\bibnamefont {Overgaag}}, \bibinfo {author}
  {\bibfnamefont {D.}~\bibnamefont {Vanmaekelbergh}}, \ and\ \bibinfo {author}
  {\bibfnamefont {W.~L.}\ \bibnamefont {Vos}},\ }\href {\doibase
  10.1038/nature02772} {\bibfield  {journal} {\bibinfo  {journal} {Nature}\
  }\textbf {\bibinfo {volume} {430}},\ \bibinfo {pages} {654} (\bibinfo {year}
  {2004})}\BibitemShut {NoStop}%
\bibitem [{\citenamefont {Lodahl}\ \emph {et~al.}(2015)\citenamefont {Lodahl},
  \citenamefont {Mahmoodian},\ and\ \citenamefont {Stobbe}}]{Lodahl2015}%
  \BibitemOpen
  \bibfield  {author} {\bibinfo {author} {\bibfnamefont {P.}~\bibnamefont
  {Lodahl}}, \bibinfo {author} {\bibfnamefont {S.}~\bibnamefont {Mahmoodian}},
  \ and\ \bibinfo {author} {\bibfnamefont {S.}~\bibnamefont {Stobbe}},\ }\href
  {\doibase 10.1103/RevModPhys.87.347} {\bibfield  {journal} {\bibinfo
  {journal} {Reviews of Modern Physics}\ }\textbf {\bibinfo {volume} {87}},\
  \bibinfo {pages} {347} (\bibinfo {year} {2015})}\BibitemShut {NoStop}%
\bibitem [{\citenamefont {Sipahigil}\ \emph {et~al.}(2016)\citenamefont
  {Sipahigil}, \citenamefont {Evans}, \citenamefont {Sukachev}, \citenamefont
  {Burek}, \citenamefont {Borregaard}, \citenamefont {Bhaskar}, \citenamefont
  {Nguyen}, \citenamefont {Pacheco}, \citenamefont {Atikian}, \citenamefont
  {Meuwly}, \citenamefont {Camacho}, \citenamefont {Jelezko}, \citenamefont
  {Bielejec}, \citenamefont {Park}, \citenamefont {Lon{\v{c}}ar},\ and\
  \citenamefont {Lukin}}]{Sipahigil2016}%
  \BibitemOpen
  \bibfield  {author} {\bibinfo {author} {\bibfnamefont {A.}~\bibnamefont
  {Sipahigil}}, \bibinfo {author} {\bibfnamefont {R.~E.}\ \bibnamefont
  {Evans}}, \bibinfo {author} {\bibfnamefont {D.~D.}\ \bibnamefont {Sukachev}},
  \bibinfo {author} {\bibfnamefont {M.~J.}\ \bibnamefont {Burek}}, \bibinfo
  {author} {\bibfnamefont {J.}~\bibnamefont {Borregaard}}, \bibinfo {author}
  {\bibfnamefont {M.~K.}\ \bibnamefont {Bhaskar}}, \bibinfo {author}
  {\bibfnamefont {C.~T.}\ \bibnamefont {Nguyen}}, \bibinfo {author}
  {\bibfnamefont {J.~L.}\ \bibnamefont {Pacheco}}, \bibinfo {author}
  {\bibfnamefont {H.~A.}\ \bibnamefont {Atikian}}, \bibinfo {author}
  {\bibfnamefont {C.}~\bibnamefont {Meuwly}}, \bibinfo {author} {\bibfnamefont
  {R.~M.}\ \bibnamefont {Camacho}}, \bibinfo {author} {\bibfnamefont
  {F.}~\bibnamefont {Jelezko}}, \bibinfo {author} {\bibfnamefont
  {E.}~\bibnamefont {Bielejec}}, \bibinfo {author} {\bibfnamefont
  {H.}~\bibnamefont {Park}}, \bibinfo {author} {\bibfnamefont {M.}~\bibnamefont
  {Lon{\v{c}}ar}}, \ and\ \bibinfo {author} {\bibfnamefont {M.~D.}\
  \bibnamefont {Lukin}},\ }\href {\doibase 10.1126/science.aah6875} {\bibfield
  {journal} {\bibinfo  {journal} {Science}\ }\textbf {\bibinfo {volume}
  {354}},\ \bibinfo {pages} {847} (\bibinfo {year} {2016})}\BibitemShut
  {NoStop}%
\bibitem [{\citenamefont {van Loo}\ \emph {et~al.}(2013)\citenamefont {van
  Loo}, \citenamefont {Fedorov}, \citenamefont {Lalumiere}, \citenamefont
  {Sanders}, \citenamefont {Blais},\ and\ \citenamefont
  {Wallraff}}]{VanLoo2013}%
  \BibitemOpen
  \bibfield  {author} {\bibinfo {author} {\bibfnamefont {A.~F.}\ \bibnamefont
  {van Loo}}, \bibinfo {author} {\bibfnamefont {A.}~\bibnamefont {Fedorov}},
  \bibinfo {author} {\bibfnamefont {K.}~\bibnamefont {Lalumiere}}, \bibinfo
  {author} {\bibfnamefont {B.~C.}\ \bibnamefont {Sanders}}, \bibinfo {author}
  {\bibfnamefont {A.}~\bibnamefont {Blais}}, \ and\ \bibinfo {author}
  {\bibfnamefont {A.}~\bibnamefont {Wallraff}},\ }\href {\doibase
  10.1126/science.1244324} {\bibfield  {journal} {\bibinfo  {journal}
  {Science}\ }\textbf {\bibinfo {volume} {342}},\ \bibinfo {pages} {1494}
  (\bibinfo {year} {2013})}\BibitemShut {NoStop}%
\bibitem [{\citenamefont {Mirhosseini}\ \emph {et~al.}(2019)\citenamefont
  {Mirhosseini}, \citenamefont {Kim}, \citenamefont {Zhang}, \citenamefont
  {Sipahigil}, \citenamefont {Dieterle}, \citenamefont {Keller}, \citenamefont
  {Asenjo-Garcia}, \citenamefont {Chang},\ and\ \citenamefont
  {Painter}}]{Mirhosseini2019}%
  \BibitemOpen
  \bibfield  {author} {\bibinfo {author} {\bibfnamefont {M.}~\bibnamefont
  {Mirhosseini}}, \bibinfo {author} {\bibfnamefont {E.}~\bibnamefont {Kim}},
  \bibinfo {author} {\bibfnamefont {X.}~\bibnamefont {Zhang}}, \bibinfo
  {author} {\bibfnamefont {A.}~\bibnamefont {Sipahigil}}, \bibinfo {author}
  {\bibfnamefont {P.~B.}\ \bibnamefont {Dieterle}}, \bibinfo {author}
  {\bibfnamefont {A.~J.}\ \bibnamefont {Keller}}, \bibinfo {author}
  {\bibfnamefont {A.}~\bibnamefont {Asenjo-Garcia}}, \bibinfo {author}
  {\bibfnamefont {D.~E.}\ \bibnamefont {Chang}}, \ and\ \bibinfo {author}
  {\bibfnamefont {O.}~\bibnamefont {Painter}},\ }\href {\doibase
  10.1038/s41586-019-1196-1} {\bibfield  {journal} {\bibinfo  {journal}
  {Nature}\ }\textbf {\bibinfo {volume} {569}},\ \bibinfo {pages} {692}
  (\bibinfo {year} {2019})}\BibitemShut {NoStop}%
\bibitem [{\citenamefont {Kannan}\ \emph {et~al.}(2020)\citenamefont {Kannan},
  \citenamefont {Ruckriegel}, \citenamefont {Campbell}, \citenamefont {{Frisk
  Kockum}}, \citenamefont {Braum{\"{u}}ller}, \citenamefont {Kim},
  \citenamefont {Kjaergaard}, \citenamefont {Krantz}, \citenamefont {Melville},
  \citenamefont {Niedzielski}, \citenamefont {Veps{\"{a}}l{\"{a}}inen},
  \citenamefont {Winik}, \citenamefont {Yoder}, \citenamefont {Nori},
  \citenamefont {Orlando}, \citenamefont {Gustavsson},\ and\ \citenamefont
  {Oliver}}]{Kannan2020}%
  \BibitemOpen
  \bibfield  {author} {\bibinfo {author} {\bibfnamefont {B.}~\bibnamefont
  {Kannan}}, \bibinfo {author} {\bibfnamefont {M.~J.}\ \bibnamefont
  {Ruckriegel}}, \bibinfo {author} {\bibfnamefont {D.~L.}\ \bibnamefont
  {Campbell}}, \bibinfo {author} {\bibfnamefont {A.}~\bibnamefont {{Frisk
  Kockum}}}, \bibinfo {author} {\bibfnamefont {J.}~\bibnamefont
  {Braum{\"{u}}ller}}, \bibinfo {author} {\bibfnamefont {D.~K.}\ \bibnamefont
  {Kim}}, \bibinfo {author} {\bibfnamefont {M.}~\bibnamefont {Kjaergaard}},
  \bibinfo {author} {\bibfnamefont {P.}~\bibnamefont {Krantz}}, \bibinfo
  {author} {\bibfnamefont {A.}~\bibnamefont {Melville}}, \bibinfo {author}
  {\bibfnamefont {B.~M.}\ \bibnamefont {Niedzielski}}, \bibinfo {author}
  {\bibfnamefont {A.}~\bibnamefont {Veps{\"{a}}l{\"{a}}inen}}, \bibinfo
  {author} {\bibfnamefont {R.}~\bibnamefont {Winik}}, \bibinfo {author}
  {\bibfnamefont {J.~L.}\ \bibnamefont {Yoder}}, \bibinfo {author}
  {\bibfnamefont {F.}~\bibnamefont {Nori}}, \bibinfo {author} {\bibfnamefont
  {T.~P.}\ \bibnamefont {Orlando}}, \bibinfo {author} {\bibfnamefont
  {S.}~\bibnamefont {Gustavsson}}, \ and\ \bibinfo {author} {\bibfnamefont
  {W.~D.}\ \bibnamefont {Oliver}},\ }\href {\doibase 10.1038/s41586-020-2529-9}
  {\bibfield  {journal} {\bibinfo  {journal} {Nature}\ }\textbf {\bibinfo
  {volume} {583}},\ \bibinfo {pages} {775} (\bibinfo {year}
  {2020})}\BibitemShut {NoStop}%
\bibitem [{\citenamefont {Luo}\ \emph {et~al.}(2019)\citenamefont {Luo},
  \citenamefont {Chen}, \citenamefont {Zhang}, \citenamefont {Zhang},
  \citenamefont {Yu}, \citenamefont {Kong}, \citenamefont {Tian}, \citenamefont
  {Zhang}, \citenamefont {Shan}, \citenamefont {Luo}, \citenamefont {Yang},
  \citenamefont {Sandoghdar}, \citenamefont {Dong},\ and\ \citenamefont
  {Hou}}]{Luo2019}%
  \BibitemOpen
  \bibfield  {author} {\bibinfo {author} {\bibfnamefont {Y.}~\bibnamefont
  {Luo}}, \bibinfo {author} {\bibfnamefont {G.}~\bibnamefont {Chen}}, \bibinfo
  {author} {\bibfnamefont {Y.}~\bibnamefont {Zhang}}, \bibinfo {author}
  {\bibfnamefont {L.}~\bibnamefont {Zhang}}, \bibinfo {author} {\bibfnamefont
  {Y.}~\bibnamefont {Yu}}, \bibinfo {author} {\bibfnamefont {F.}~\bibnamefont
  {Kong}}, \bibinfo {author} {\bibfnamefont {X.}~\bibnamefont {Tian}}, \bibinfo
  {author} {\bibfnamefont {Y.}~\bibnamefont {Zhang}}, \bibinfo {author}
  {\bibfnamefont {C.}~\bibnamefont {Shan}}, \bibinfo {author} {\bibfnamefont
  {Y.}~\bibnamefont {Luo}}, \bibinfo {author} {\bibfnamefont {J.}~\bibnamefont
  {Yang}}, \bibinfo {author} {\bibfnamefont {V.}~\bibnamefont {Sandoghdar}},
  \bibinfo {author} {\bibfnamefont {Z.}~\bibnamefont {Dong}}, \ and\ \bibinfo
  {author} {\bibfnamefont {J.~G.}\ \bibnamefont {Hou}},\ }\href {\doibase
  10.1103/PhysRevLett.122.233901} {\bibfield  {journal} {\bibinfo  {journal}
  {Physical Review Letters}\ }\textbf {\bibinfo {volume} {122}},\ \bibinfo
  {pages} {233901} (\bibinfo {year} {2019})}\BibitemShut {NoStop}%
\bibitem [{\citenamefont {Stiesdal}\ \emph {et~al.}(2020)\citenamefont
  {Stiesdal}, \citenamefont {Busche}, \citenamefont {Kumlin}, \citenamefont
  {Kleinbeck}, \citenamefont {B{\"{u}}chler},\ and\ \citenamefont
  {Hofferberth}}]{Stiesdal2020}%
  \BibitemOpen
  \bibfield  {author} {\bibinfo {author} {\bibfnamefont {N.}~\bibnamefont
  {Stiesdal}}, \bibinfo {author} {\bibfnamefont {H.}~\bibnamefont {Busche}},
  \bibinfo {author} {\bibfnamefont {J.}~\bibnamefont {Kumlin}}, \bibinfo
  {author} {\bibfnamefont {K.}~\bibnamefont {Kleinbeck}}, \bibinfo {author}
  {\bibfnamefont {H.~P.}\ \bibnamefont {B{\"{u}}chler}}, \ and\ \bibinfo
  {author} {\bibfnamefont {S.}~\bibnamefont {Hofferberth}},\ }\href
  {http://arxiv.org/abs/2005.05089} {\  (\bibinfo {year} {2020})},\ \Eprint
  {http://arxiv.org/abs/2005.05089} {arXiv:2005.05089} \BibitemShut {NoStop}%
\bibitem [{\citenamefont {Mollow}(1975)}]{Mollow1975}%
  \BibitemOpen
  \bibfield  {author} {\bibinfo {author} {\bibfnamefont {B.~R.}\ \bibnamefont
  {Mollow}},\ }\href {\doibase 10.1103/PhysRevA.12.1919} {\bibfield  {journal}
  {\bibinfo  {journal} {Physical Review A}\ }\textbf {\bibinfo {volume} {12}},\
  \bibinfo {pages} {1919} (\bibinfo {year} {1975})}\BibitemShut {NoStop}%
\bibitem [{\citenamefont {Gardiner}\ and\ \citenamefont
  {Zoller}(2004)}]{Gardiner2004}%
  \BibitemOpen
  \bibfield  {author} {\bibinfo {author} {\bibfnamefont {C.~W.}\ \bibnamefont
  {Gardiner}}\ and\ \bibinfo {author} {\bibfnamefont {P.}~\bibnamefont
  {Zoller}},\ }\href@noop {} {\emph {\bibinfo {title} {{Quantum noise : a
  handbook of Markovian and non-Markovian quantum stochastic methods with
  applications to quantum optics}}}}\ (\bibinfo  {publisher} {Springer},\
  \bibinfo {year} {2004})\ p.\ \bibinfo {pages} {449}\BibitemShut {NoStop}%
\bibitem [{\citenamefont {Yudson}(1985)}]{Yudson1985}%
  \BibitemOpen
  \bibfield  {author} {\bibinfo {author} {\bibfnamefont {V.}~\bibnamefont
  {Yudson}},\ }\href@noop {} {\bibfield  {journal} {\bibinfo  {journal} {Zh.
  Eksp. Teor. Fiz}\ }\textbf {\bibinfo {volume} {88}},\ \bibinfo {pages} {1757}
  (\bibinfo {year} {1985})}\BibitemShut {NoStop}%
\end{thebibliography}%

\end{document}